\documentstyle[12pt,eqsection]{article}
\newcommand{\be}{\begin{equation}}
\newcommand{\ee}{\end{equation}}
\newcommand{\bea}{\begin{eqnarray}}
\newcommand{\eea}{\end{eqnarray}}
\newcommand{\norsl}{\normalsize\sl}
\newcommand{\norsc}{\normalsize\sc}

\def\Av{\mbox{\boldmath $A$}}

\def\Cv{\mbox{\boldmath $C$}}

\def\Kv{\mbox{\boldmath $K$}}

\def\qv{\mbox{\boldmath $q$}}

\begin{document}

\begin{titlepage}

\title{Factorization Scheme and Parton Distributions\\
in the Polarized Virtual Photon Target
}
\author{
\norsc  Ken SASAKI\thanks{e-mail address: sasaki@phys.ynu.ac.jp}~ and
       Tsuneo UEMATSU\thanks{e-mail address: uematsu@phys.h.kyoto-u.ac.jp} \\
\norsl  Dept. of Physics,  Faculty of Engineering, Yokohama National
University \\
\norsl  Yokohama 240-8501, JAPAN \\
\norsl  Dept. of Fundamental Sciences, FIHS, Kyoto University \\
\norsl     Kyoto 606-8501, JAPAN \\
}

\date{}
\maketitle

\begin{abstract}
{\normalsize
We investigate spin-dependent parton distributions in the polarized virtual
photon target in perturbative QCD up to the next-to-leading order (NLO).
In the case $\Lambda^2 \ll P^2 \ll Q^2$, where $-Q^2$ ($-P^2$) is the mass
squared of the probe (target) photon, parton distributions can be predicted
completely up to NLO, but they are factorization-scheme-dependent.
We analyze parton distributions in six different factorization schemes
and discuss their scheme dependence. We study, in particular, the
QCD and QED axial anomaly
effects on the first moments of parton distributions to see
the interplay between the axial anomalies and factorization schemes.
We also show that the factorization-scheme dependence is characterized
by the large-$x$ behaviors of quark distributions in the virtual photon.
Gluon distribution is predicted to be the same up to NLO among the six
factorization schemes examined. In particular,
the first moment of gluon distribution is found to be  factorization-scheme
independent up to NLO.
}
\end{abstract}

\begin{picture}(5,2)(-290,-500)
\put(2.3,0){YNU-HEPTh-00-102}
\put(2.3,-15){KUCP-160}
\put(2.3,-30){Revised}
\put(2.3,-45){December 2000}
\end{picture}

\thispagestyle{empty}
\end{titlepage}
\setcounter{page}{1}
\baselineskip 18pt
\section{Introduction}
\smallskip

In the two-photon process of $e^+ e^-$ collision experiments,
we can measure the structure functions of the  virtual photon (Fig.1).
The advantage in  studying the virtual photon target is that,
in the case
\begin{equation}
\Lambda^2 \ll P^2 \ll Q^2
\label{kin}
\end{equation}
where $-Q^2$ ($-P^2$) is the mass squared of
the probe (target) photon,
and $\Lambda$ is the QCD scale parameter, we can
calculate the whole structure function up to the next-to-leading order (NLO) in
QCD by the perturbative method, in contrast to the case  of the real
photon target
where in  NLO there exist non-perturbative pieces~\cite{BB,DO}.
The spin-independent structure functions $F_2^{\gamma}(x, Q^2, P^2)$
and $F_L^{\gamma}(x, Q^2, P^2)$ as well as the parton contents
were studied in the leading order (LO)
\cite{UW1} and in NLO \cite{UW2}-\cite{SS}. The target mass effect
of unpolarized and polarized virtual photon structure in LO was discussed
in Ref.\cite{MR}.

The information on the spin structure of the photon would be provided by the
resolved photon process in polarized version of the DESY electron and proton
collider HERA  \cite{Barber,SVZ}.
More directly, polarized photon structure function can
be measured by the polarized $e^+e^-$ collision in the future linear
colliders.
For the real photon ($P^2=0$) target, there exists only one
spin-dependent struture function,
$g_1^\gamma(x,Q^2)$, which is equivalent to the structure function $W_4^\gamma
(x,Q^2)$  ($g_1^\gamma\equiv 2W_4^\gamma$) discussed some time
ago in \cite{BM, AMR}.  The LO QCD corrections to
$g_1^\gamma$ for the real photon
target was first calculated by one of the authors~\cite{KS1} and
later in Refs.\cite{MANO,BASS}, while
the NLO QCD analysis was performed by Stratmann and Vogelsang \cite{SV}.
The first moment of the photon structure function $g_1^\gamma$ has recently
attracted attention in the literature \cite{BASS,ET0,NSV,FS,BBS} in connection
with its relevance for the axial anomaly.
More recently the present authors investigated~\cite{SU} the spin-dependent
structure function $g_1^{\gamma}(x, Q^2, P^2)$ of the  virtual photon up to
NLO in QCD, where $P^2$ is in the above kinematical region (\ref{kin}).
The analysis was made  in the
framework of the operator product expansion (OPE) supplemented by the
renormalization group method and also in the framework of the QCD improved
parton model~\cite{A} using the DGLAP parton evolution equations.


In the past few years, accuracy of the experimental data on the
spin-dependent structure function $g_1$ of the nucleon has been
significantly improved~\cite{Nucleon}.
Using these experimental data together with the already existing world data,
several  groups~\cite{AnalysisGRSVGS}-\cite{AnalysisEX} have
carried out the NLO QCD analyses on the  polarized parton
distributions in the nucleon.
These parton distributions may be used for predicting the behaviors of other
processes such as polarized Drell-Yan reactions and polarized semi-inclusive
deep inelastic scatterings, and etc.
However, parton distributions obtained from the NLO analyses
are dependent on the factorization scheme employed.
It is possible that  parton distributions obtained in one scheme
may be more appropriate to use than those in other schemes.
In the case of  nucleon target, however,  it may be difficult
to examine the features of each factorization scheme, since for the moment
it is inevitable to resort to some assumptions in order to extract parton
distributions from the experimental data.

On the other hand, it is remarkable that, in the case of
virtual photon target with a virtual mass $-P^2$ being
in the kinematical  region of Eq.(\ref{kin}),
not only the photon structure functions but also the parton distributions in
the
target  can be  predicted entirely up to NLO in QCD. Thus, comparing
the parton distributions predicted by one scheme with those by other
schemes, we can easily examine the features of each factorization scheme.
In consequence, the virtual
photon target may serve as an optimal place to study the behaviors of
parton distributions and their factorization-scheme dependence.


In this paper we examine in detail the polarized parton
(i.e., quark and gluon) distributions in the virtual photon target.
The polarized parton distributions are particularly interesting due to the fact
that  they have relevance to the axial anomaly~\cite{JK}. The interplay
between the QCD axial anomaly and  factorization schemes has already been
discussed
for the spin-dependent structure function $g_1$ of the
nucleon~\cite{AR}-\cite{HYC}.
It was explained there that the QCD axial anomaly effect is retained in
the flavor-singlet quark distribution in the nucleon in the standard
$\overline{\rm MS}$ scheme, but it is shifted to the gluon coefficient
function in such a scheme called chirally-invariant (CI)
factorization scheme.
Now it should be pointed out that the polarized photon target is unique in
the sense
that not only QCD  but also the OED axial anomaly takes place.
The QED axial anomaly, which is U(1) anomaly, emerges when a quark has an
electromagnetic charge. Thus the flavor-non-singlet quark distribution
is also relevant, besides the flavor-singlet one. Depending upon
factorization schemes, the QED axial anomaly effect resides in
both  the flavor-singlet and non-singlet polarized quark distributions
in the  virtual photon, or it is shifted to the photon coefficient
function, in which case we arrive at an interesting result: the
first  moments of the polarized quark distributions
in the virtual photon, both flavor singlet
and  non-singlet, vanish in NLO.
Also we find that the large $x$-behaviors of polarized quark distributons
dramatically vary from one factorization scheme to another. Indeed, for
$x\rightarrow 1$,  the quark distributions positively diverge or
negatively diverge or remain finite,  depending on factorization schemes.

We perform our analyses
in six different factorization schemes, (i) $\overline{\rm MS}$, (ii) CI
(chirally invariant) (it is also called as JET)~\cite{HYC,MT},   (iii) AB
(Adler-Bardeen)~\cite{BFR},  (iv) OS (off-shell)~\cite{BFR}, (v) AR
(Altarelli-Ross)~\cite{BFR}, and  finally (vi) DIS$_{\gamma}$
schemes~\cite{GRV}, and see  how the parton distributions change in each
scheme.
In particular, we study the axial anomaly effects on the first moments and the
large-$x$ behaviors of  parton distributions.
Gluon distribution in the
virtual photon is found to be the same up to NLO, at least among the
factorization schemes  considered in this paper.
Furthermore, the first moment of gluon distribution turns out to be
factorization-scheme independent up to NLO.
Part of the result has been briefly reported  elsewhere~\cite{SU2}.

In the next section we discuss
the polarized parton distributions in the virtual photon.
The explicit expressions for the flavor singlet-(non-singlet-)quark and gluon
distributions predicted in QCD up to NLO are given in Appendix A.
In Sec. 3, we derive the transformation rules for the relevant two-loop
anomalous  dimensions and one-loop photon matrix elements from the
$\overline{\rm MS}$ scheme to other factorization schemes and
then explain particular factorization schemes we consider in this paper.
In Sec. 4, we examine the first moments of parton distributions
with emphasis on the interplay between the QCD and QED axial anomalies and
the facorization schemes. The behaviors of parton distributions near $x=1$
and their factorization-scheme dependence are discussed in Sec. 5.
The numerical analyses of parton distributions
predicted by different factorization schemes  will be given in Sec. 6. The
final section is devoted to the conclusion and discussion.


\section{Polarized parton distributions in photon}
\smallskip

Let $q^i_{\pm}(x,Q^2,P^2)$, $G^{\gamma}_{\pm}(x,Q^2,P^2)$,
$\Gamma^{\gamma}_{\pm}(x,Q^2,P^2)$ be  quark with
$i$-flavor, gluon, and  photon distribution functions with $\pm$ helicities
of the longitudinally polarized virtual photon with mass $-P^2$.
Then the spin-dependent parton distributions are defined as
$\Delta q^i \equiv q^i_+ + \bar{q}^i_+ -q^i_- -\bar{q}^i_-$~,
$\Delta G^{\gamma} \equiv G^{\gamma}_+ -G^{\gamma}_- $~, and
$\Delta \Gamma^{\gamma} \equiv \Gamma^{\gamma}_+ -\Gamma^{\gamma}_-$~.
In the leading order of the electromagnetic coupling constant,
$\alpha=e^2/4\pi$,
$\Delta \Gamma^{\gamma}$ does not evolve with $Q^2$ and is set to be
$\Delta \Gamma^{\gamma}(x,Q^2,P^2)=\delta(1-x)$. For later convenience we use,
instead of $\Delta q^i$, the flavor singlet and non-singlet combinations of
spin-dependent quark distributions as follows:
\bea
      \Delta q^{\gamma}_S &\equiv& \sum_i~ \Delta q^i~,  \nonumber \\
    \Delta q^{\gamma}_{NS} &\equiv& \sum_i e_i^2 \Bigl( \Delta q^i -
\frac{\Delta
q^{\gamma}_S}{N_f} \Bigr)~.
\eea
In terms of these parton distributions,
the polarized virtual photon  structure function
$g_1^{\gamma}(x, Q^2, P^2)$ is expressed in the QCD improved parton model
as~\cite{SU}
\bea
  g_1^{\gamma}(x, Q^2,P^2)&=& \int^1_x \frac{dy}{y}~\biggl\{
    \Delta q^{\gamma}_S (y,Q^2,P^2)~\Delta C^{\gamma}_S (\frac{x}{y}, Q^2)
     + \Delta G^{\gamma} (y,Q^2,P^2)~\Delta C^{\gamma}_G(\frac{x}{y},Q^2)~
\nonumber  \\
  & &
\quad + \Delta q^{\gamma}_{NS}
(y,Q^2,P^2) ~\Delta C^{\gamma}_{NS} (\frac{x}{y}, Q^2)
\biggr\}   + \Delta C^{\gamma}_{\gamma} (x, Q^2) \label{Solg}~.
\eea
where $\Delta C^{\gamma}_S$($\Delta C^{\gamma}_{NS}$), $\Delta
C^{\gamma}_G$, and
$\Delta C^{\gamma}_{\gamma}$ are the coefficient functions  corresponding to
singlet(non-singlet)-quark, gluon, and photon, respectively,  and are
independent of
$P^2$. The Mellin moments of $g_1^{\gamma}$ is written  as
\be
  g_1^{\gamma}(n,Q^2,P^2)=\Delta \widetilde{\Cv}^{\gamma}(n,Q^2) \cdot
\Delta \widetilde{\qv}^{\gamma}(n,Q^2,P^2)~,
\label{gonegamma} \\
\ee
where
\bea
    \Delta \widetilde{\Cv}^{\gamma}(n,Q^2)&=&(\Delta C^{\gamma}_S~,~
  \Delta C^{\gamma}_G~,~\Delta C^{\gamma}_{NS}~,~
   \Delta C^{\gamma}_{\gamma})~, \nonumber  \\
\Delta \widetilde{\qv}^{\gamma}(n,Q^2,P^2)&=&(\Delta q_S^{\gamma}~,~
\Delta G^{\gamma}~,~ \Delta q_{NS}^{\gamma}~, ~\Delta \Gamma^{\gamma})~,
\nonumber
\eea
and the matrix notation is implicit.

The explicit expressions of $\Delta q^\gamma_S$,
$\Delta G^\gamma$, and $\Delta q^\gamma_{NS}$ up to NLO can be derived
from Eq.(4.46) of Ref.\cite{SU}, which are given in Appendix A.
They are written~\cite{Footnote1} in
terms of  one-(two-) loop hadronic anomalous dimensions
$\Delta\gamma_{ij}^{0,n}$
($\Delta\gamma_{ij}^{(1),n}$) ($i,j=\psi, G$) and
$\Delta\gamma_{NS}^{0,n}$
($\gamma_{NS}^{(1),n}$), one-(two-) loop anomalous dimensions
$\Delta K_i^{0,n}$ ($\Delta K_i^{(1),n}$) ($i=\psi, G, NS$) which
represent the
mixing between photon and three hadronic operators $R_i^n$ ($i=\psi, G,
NS$), and
finally
$\Delta A_i^n$, the one-loop photon matrix elements  of  hadronic operators
renormalized at
$\mu^2= P^2(=-p^2)$,
\be
   \langle \gamma (p) \mid R_i^n (\mu) \mid \gamma (p) \rangle \vert_{\mu^2=
P^2}
=\frac{\alpha}{4\pi}  \Delta A_i^n     \qquad  (i=\psi, G, NS)~.
\label{Initial}
\ee

The photon matrix elements $\Delta A_i^n$ are scheme-dependent.
In one-loop order, they are given, in the $\overline {\rm MS}$ scheme, by
\cite{BQ}
\bea
\Delta A_{\psi,~\overline {\rm MS}}^n &=&\frac{\langle e^2 \rangle}
{\langle e^4 \rangle -\langle e^2 \rangle^2}~
\Delta A_{NS,~\overline {\rm MS}}^n   \nonumber  \\
&=&12 \langle e^2 \rangle N_f \left\{
\frac{n-1}{n(n+1)}S_1(n)+\frac{4}{(n+1)^2} -\frac{1}{n^2}
-\frac{1}{n} \right\} \label{PhotonMSn} ~,\\
\Delta A_{G,~\overline {\rm MS}}^n &=&0~, \nonumber
\eea
where $S_1(n)=\sum^n_{j=1}\frac{1}{j}$~.

\section{Factorization schemes}
\smallskip

\subsection{Transformation rules from $\overline {\rm MS}$ scheme to
$a$-scheme}
Although $g_1^{\gamma}$ is a
physical quantity and thus unique, there remains a freedom in the
factorization of
$g_1^{\gamma}$  into $\Delta \widetilde{\Cv}^{\gamma}$ and
$\Delta \widetilde{\qv}^{\gamma}$.
Given the formula Eq.(\ref{gonegamma}),
we can always redefine $\Delta \widetilde{\Cv}^{\gamma}$ and
$\Delta \widetilde{\qv}^{\gamma}$ as follows \cite{FP1}:
\bea
  \Delta \widetilde{\Cv}^{\gamma}(n,Q^2)\rightarrow
\Delta{ \widetilde{\Cv}}^{\gamma}(n,Q^2)\vert_a
       &\equiv& \Delta \widetilde{\Cv}^{\gamma}(n,Q^2)Z^{-1}_a(n,Q^2)~,
\label{aCoeff}  \\
  \Delta \widetilde{\qv}^{\gamma}(n,Q^2,P^2)  \rightarrow
\Delta{ \widetilde{\qv}}(n,Q^2,P^2)\vert_a
  &\equiv& Z_a(n,Q^2)~ \Delta \widetilde{\qv}^{\gamma}(n,Q^2,P^2)~, \label{aPDF}
\eea
where $\Delta{\widetilde{\Cv}}^{\gamma}\vert_a  $ and $ \Delta
{ \widetilde{\qv}}\vert_a $  correspond to the quantities in a new
factorization
scheme-$a$.  Note that the coefficient functions and anomalous
dimensions are closely connected under factorization. We will study
the factorization scheme dependence of parton distribution up to  NLO, by
which we mean that a scheme transformation  for the coefficient functions is
considered up to the one-loop order, since a NLO prediction for
$g_1^{\gamma}$ is given by the one-loop coefficient functions and
anomalous dimensions up to the two-loop order.

The most general form of a transformation for the coefficient functions
in one-loop order, from the $\overline {\rm MS}$ scheme to
a new factorization scheme-$a$, is given by
\bea
     \Delta C_{S,~a}^{\gamma,~ n}&=&
\Delta C_{S,~\overline{\rm MS}}^{\gamma,~ n}~
-\langle e^2 \rangle\frac{\alpha_s}{2\pi}~\Delta w_S(n,a)~, \nonumber \\
\Delta C_{G,~a}^{\gamma,~ n}&=&\Delta C_{G,~\overline {\rm MS}}^{\gamma,~ n}~
-\langle e^2 \rangle\frac{\alpha_s}{2\pi}~\Delta z(n,a) ~, \nonumber\\
\Delta C_{NS,~a}^{\gamma,~ n}&=&\Delta C_{NS,~\overline {\rm MS}}^{\gamma,~ n}~
-\frac{\alpha_s}{2\pi}~\Delta w_{NS}(n,a)~, \label{Coeffgamma} \\
  \Delta C_{\gamma,~a}^{\gamma,~ n}&=&
\Delta C_{\gamma,~\overline{\rm MS}}^{\gamma,~ n}~
-\frac{\alpha}{\pi}~3\langle e^4 \rangle \Delta {\hat z}(n,a)~, \nonumber
\eea
where $\langle e^4 \rangle =\sum_i e^4_i/N_f$.
The flavor-singlet(nonsinglet) quark coefficient functions are expanded up to
the one-loop order as
\bea
  \Delta C_S^{\gamma,~n}&=&\langle e^2
\rangle\Bigl(1+\frac{\alpha_s}{4\pi}\Delta
B_S^n +{\cal O}
(\alpha_s^2)~\Bigr) ~,\\
\Delta C_{NS}^{\gamma,~n}&=&1+\frac{\alpha_s}{4\pi}\Delta B_{NS}^n
+{\cal O} (\alpha_s^2)~,\nonumber
\eea
with $\Delta B_S^n=\Delta B_{NS}^n$~.
The $\Delta z(n,a)$ ($\Delta {\hat z}(n,a)$) term tells how much of the
QCD (QED) axial anomaly effect is transferred to the coefficient function
in the new factorization scheme.
The gluon and photon coefficient functions $\Delta C_{G}^{\gamma,~ n}$ and
$\Delta C_{\gamma}^{\gamma,~ n}$ start from the one-loop order (i.e.,
from the NLO):
\bea
  \Delta C_G^{\gamma,~n}&=&\langle e^2 \rangle\Bigl(\frac{\alpha_s}{4\pi}\Delta
B_G^n +{\cal O}
(\alpha_s^2)~\Bigr)~, \nonumber\\
  \Delta C^{\gamma,~n}_\gamma&=&\frac{\alpha}{4\pi}3N_f\langle e^4 \rangle
      \Bigl(\Delta B_{\gamma}^n +{\cal O}(\alpha_s)~\Bigr)~.
\eea
In the $\overline {\rm MS}$ scheme,
$\Delta C_{\gamma,~\overline {\rm MS}}^{\gamma,~ n}$ has been obtained  from
$\Delta C_{G,~\overline {\rm MS}}^{\gamma,~ n}$,  with changes:
$\alpha_s/{2\pi}\rightarrow
(2\alpha/{\alpha_s})\times (\alpha_s/{2\pi})$~,
$\langle e^2 \rangle\rightarrow 3\langle e^4 \rangle $,  and $3$ is the
number of
colors. Thus we have
\be
     \Delta B_{\gamma,~\overline {\rm MS}}^n=\frac{2}{N_f}
\Delta B_{G,~\overline {\rm MS}}^n~. \label{GgammaMS}
\ee
Since, in the leading order, coefficient functions are given by
\be
\Delta\Cv^{\gamma}_{\overline{\rm MS}}\vert_{\rm LO}=
\Delta\Cv^{\gamma}_a\vert_{\rm LO}=(\langle e^2 \rangle,0,1,0)~,
\ee
the relations (\ref{Coeffgamma}) between the coefficient functions in the
$a$-scheme and $\overline {\rm MS}$ scheme lead to
$Z^{-1}_a(n,Q^2)$, which is expressed as
\bea
   && Z^{-1}_a(n,Q^2) \nonumber  \\
&& \nonumber  \\
&=&I -  \left(\matrix{\frac{\alpha_s}{2\pi}\Delta w_S(n,a) &
\frac{\alpha_s}{2\pi} \Delta z(n,a) &0 &  \frac{\alpha}{\pi}3
\langle e^2 \rangle \Delta {\hat z}(n,a) \cr
               0 & & 0&0\cr
   0&0&\frac{\alpha_s}{2\pi}\Delta w_{NS}(n,a)&
\frac{\alpha}{\pi}3(\langle e^4 \rangle -\langle e^2 \rangle^2)\Delta {\hat
z}(n,a) \cr 0&0&0&0}
\right)~,
\nonumber  \\ \label{Zinverse}
\eea
where $I$ is a $4\times 4$ unit matrix.

Now we derive corresponding
transformation rules from $\overline {\rm MS}$ scheme to $a$-scheme for the
relevant two-loop anomalous dimensions.
The parton distribution functions $\Delta \widetilde{\qv}^{\gamma}(n,Q^2,P^2)$
satisfy  the following evolution equation \cite{GRV,FP1,GR,FP2,AGF}:
\be
\frac{d\Delta \widetilde{\qv}^{\gamma}(n,Q^2,P^2)}{d~{\rm ln}Q^2}=
\Delta \widetilde{P}(n,Q^2)\Delta \widetilde{\qv}^{\gamma}(n,Q^2,P^2)~,
\ee
where
\be
\Delta \widetilde{P}(n,Q^2) =  \left(\matrix{\Delta P_{\psi\psi}(n,Q^2) &
\Delta P_{\psi G}(n,Q^2) &0 & \Delta k_S(n,Q^2)\cr
\Delta P_{G \psi}(n,Q^2) & \Delta P_{G G}(n,Q^2) & 0&\Delta k_G(n,Q^2)\cr
   0&0&\Delta P_{NS}(n,Q^2)&
\Delta k_{NS}(n,Q^2) \cr 0&0&0&0} \right)~.\nonumber \label{4by4matrix}\\
\ee
\relax From Eq.(\ref{aPDF}), we obtain
\bea
\frac{d\Delta \widetilde{\qv}^{\gamma}(n,Q^2,P^2)\vert_a}{d~{\rm ln}Q^2}&=&
\frac{dZ_a(n,Q^2) }{d~{\rm ln}Q^2}\Delta \widetilde{\qv}^{\gamma}(n,Q^2,P^2)
\vert_{\overline {\rm MS}}+
Z_a(n,Q^2)\frac{d\Delta \widetilde{\qv}^{\gamma}(n,Q^2,P^2)
\vert_{\overline {\rm MS}}}{d~{\rm ln}Q^2}  \nonumber \\
&=&\Delta \widetilde{P}(n,Q^2)\vert_a~\Delta
\widetilde{\qv}^{\gamma}(n,Q^2,P^2)\vert_a~,
\eea
with
\be
\Delta \widetilde{P}(n,Q^2)\vert_a=\left[ \frac{dZ_a(n,Q^2) }{d~{\rm ln}Q^2}+
Z_a(n,Q^2) \Delta \widetilde{P}(n,Q^2)\vert_{\overline {\rm MS}}
\right]~Z^{-1}_a(n,Q^2)~.
\ee
The splitting functions $\Delta P_{i}(n,Q^2)$ ($i=\psi\psi,
\psi G, G\psi, GG$, and $NS$) and $\Delta k_j(n,Q^2)$ ($j=S,G,NS$) are
expanded as
\bea
\Delta P_{i}(n,Q^2)&=&\frac{\alpha_s(Q^2)}{2\pi}\Delta P^{(0)}_{i}(n)
+\left[\frac{\alpha_s(Q^2)}{2\pi}\right]^2 \Delta P^{(1)}_{i}(n)
+\cdots~, \\
\Delta k_j(n,Q^2)&=&\frac{\alpha}{2\pi}\Delta k^{(0)}_{j}(n)
+\frac{\alpha~\alpha_s(Q^2)}{(2\pi)^2} \Delta k^{(1)}_{j}(n)
+\cdots~,
\eea
Since the QCD effective coupling constant $\alpha_s(Q^2)$ satisfies
\be
\frac{d~\alpha_s(Q^2)}{d~{\rm ln}Q^2}=
-\beta_0\frac{\alpha_s(Q^2)^2}{4\pi}+\cdots~,
\ee
where $\beta_0=11-\frac{2}{3}N_f$ is the one-loop coefficient of the
QCD beta function, and the $n$-th anomalous dimensions are defined as
\bea
    \Delta P^{(0)}_{i}(n) &=&-\frac{1}{4}\Delta \gamma^{0,n}_i~, \qquad
    \Delta P^{(1)}_{i}(n) =-\frac{1}{8}\Delta \gamma^{(1),n}_i ~, \\
\Delta k^{(0)}_{j}(n) &=&\frac{1}{4}\Delta K^{0,n}_{j}~, \qquad
    \Delta k^{(1)}_{j}(n) =\frac{1}{8}\Delta K^{(1),n}_{j}~,
\eea
we find for one-loop
\be
    \Delta \gamma^{0,n}_{i,~a} = \Delta \gamma^{0,n}_{i,~\overline {\rm
MS}}~,
  \qquad \Delta K^{0,n}_{j,~a}=\Delta K^{0,n}_{j,~\overline {\rm MS}}~,
\ee
and for two-loop
\bea
   \Delta \gamma^{(1),n}_{\psi\psi,~a}&=&\Delta
\gamma^{(1),n}_{\psi\psi,~\overline{\rm MS}}
    + 2 \Delta z(n,a)~\Delta \gamma^{0,n}_{G\psi} +4 \beta_0 \Delta
w_S(n,a)~,\nonumber\\
  \Delta \gamma^{(1),n}_{\psi G,~a}&=&\Delta \gamma^{(1),n}_{\psi
G,~\overline {\rm MS}}
    + 2 \Delta z(n,a)~\Bigl[ \Delta \gamma^{0,n}_{GG} -
   \Delta \gamma^{0,n}_{\psi\psi} + 2\beta_0 \Bigr]  \nonumber  \\
& &+ 2\Delta w_S(n,a) \Delta \gamma^{0,n}_{\psi G}~,\nonumber\\
  \Delta \gamma^{(1),n}_{G\psi,~a}&=&\Delta \gamma^{(1),n}_{G\psi,
~\overline{\rm MS}}
- 2\Delta w_S(n,a) \Delta \gamma^{0,n}_{G\psi}~,\nonumber\\
  \Delta \gamma^{(1),n}_{GG,~a}&=&\Delta \gamma^{(1),n}_{GG,~\overline {\rm MS}}
  - 2 \Delta z(n,a) \Delta \gamma^{0,n}_{G\psi} ~,\nonumber\\
     \Delta \gamma^{(1),n}_{NS,~a}&=&\Delta
\gamma^{(1),n}_{NS,~\overline {\rm MS}}
+4 \beta_0 \Delta w_{NS}(n,a)~,\label{Transformation}\\
   \Delta K^{(1),n}_{S,~a}&=& \Delta K^{(1),n}_{S,~\overline {\rm MS}}
+  2\Delta w_S(n,a) \Delta K^{0,n}_S
+ 4 \Delta {\hat z}(n,a)3\langle e^2 \rangle \Delta \gamma^{0,n}_{\psi\psi}
~,\nonumber\\
  \Delta K^{(1),n}_{G,~a}&=& \Delta K^{(1),n}_{G,~\overline {\rm MS}}
+ 4 \Delta {\hat z}(n,a)3\langle e^2 \rangle \Delta
\gamma^{0,n}_{G\psi}~,\nonumber \\
  \Delta K^{(1),n}_{NS,~a}&=& \Delta K^{(1),n}_{NS,~\overline {\rm MS}}
  +  2\Delta w_{NS}(n,a) \Delta K^{0,n}_{NS}  \nonumber \nonumber \\
& & + 4 \Delta {\hat z}(n,a)3(\langle e^4 \rangle -\langle e^2 \rangle^2)
\Delta
\gamma^{0,n}_{NS}~.
\nonumber
\eea

The one-loop photon matrix elements of the hadronic operators,
$\Delta A^n_{\psi}$ and  $ \Delta A^n_{NS}$ in Eq.(\ref{Initial}),
are related to each other as
\be
\Delta A^n_{NS}=\Delta  A^n_{\psi}~\frac{\langle e^4 \rangle -\langle e^2
\rangle^2}{\langle e^2 \rangle}~,
\label{ChargeFactor}
\ee
and the sum
\be
(\Delta C_\gamma^{\gamma,~n}/\frac{\alpha}{4\pi}+\langle e^2 \rangle
\Delta A^n_{\psi}+\Delta A^n_{NS})
\ee
is factorization-scheme-independent in one-loop order \cite{SU}.
Thus we obtain from Eq.(\ref{Coeffgamma})
  \bea
     \Delta A_{\psi,~a}^n&=&\Delta A_{\psi,~\overline {\rm MS}}^n +12\langle e^2
\rangle\Delta {\hat z}(n,a)~,
\nonumber\\
  \Delta A_{G,~a}^n&=&\Delta A_{G,~\overline {\rm MS}}^n =0~,
\label{TransForMat}\\
  \Delta A_{NS,~a}^n&=&\Delta A_{NS,~\overline {\rm MS}}^n
+12(\langle e^4 \rangle -\langle e^2 \rangle^2)~\Delta {\hat z}(n,a)~.
\nonumber
\eea
Note that $\Delta A^n_G=0$ in one-loop order.

It is possible to choose $\Delta z(n,a)$ and $\Delta {\hat z}(n,a)$
arbitrarily.
In the following, we take $\Delta {\hat z}(n,a)=\Delta z(n,a)$ in the CI-like
schemes  and $\Delta {\hat z}(n,{\rm DIS}_{\gamma})\neq \Delta z(n,{\rm
DIS}_{\gamma})=0$  in the ${\rm DIS}_{\gamma}$  scheme.
In one-loop order  we have $\Delta w_S(n,a)=\Delta w_{NS}(n,a)$. Thus from
now on,
we set $\Delta w_S(n,a)=\Delta w_{NS}(n,a)\equiv \Delta w(n,a)$.
Let us now discuss the features of several factorization schemes.

\subsection{The $\overline {\rm MS}$ scheme}
This is the only scheme in which both relevant one-loop coefficient
functions and
two-loop anomalous dimensions were actually calculated~\cite{BQ,KMMSU,MvN,V}.
In fact there still remain ambuguities in the $\overline {\rm MS}$ scheme,
depending on
how to handle $\gamma_5$ in $n$ dimensions.
The $\overline {\rm MS}$ scheme we call here is the one due to Mertig and van
Neerven~\cite{MvN} and  Vogelsang~\cite{V}, in which the first moment of the
non-singlet
quark operator  vanishes, corresponding to the conservation of the
non-singlet axial
current. Indeed we have $\Delta \gamma^{(1),n=1}_{NS,~\overline {\rm MS}}=0$.
Explicit expressions of the relevant one-loop coefficient functions and
two-loop anomalous dimensions can be found, for example, in Appendix
of  Ref.~\cite{SU}. It is noted that,
in the  $\overline {\rm MS}$ scheme,  both the QCD and QED  axial
anomalies reside  in the quark distributions and not in the gluon
and photon coefficient functions.
In fact we observe
\bea
\Delta \gamma^{(1),n=1}_{\psi\psi,~\overline {\rm MS}}&=&
24C_FT_f \neq 0~, \\
\Delta B_{G,~\overline {\rm MS}}^{n=1}&=&\Delta B_{\gamma,~
\overline{\rm MS}}^{n=1}=0~.
\eea
where $C_F=\frac{4}{3}$ and $T_f=\frac{N_f}{2}$.
Also we find from Eq.(\ref{PhotonMSn}) that the first moments of the one-loop
photon matrix elements of quark operators gain the non-zero values, i.e.,
\be
\Delta A_{\psi,~\overline {\rm MS}}^{n=1}=\frac{\langle e^2 \rangle}{\langle
e^4
\rangle -\langle e^2
\rangle^2}~\Delta A_{NS,~\overline {\rm MS}}^{n=1}=-12\langle e^2 \rangle N_f
~,\label{PhotonMSn=1}
\ee
which is due to the QED axial anomaly.

\subsection{The CI-like schemes}

The EMC measurement~\cite{EMC} of the first moment of the proton spin structure
function  $g_1^p(x,Q^2)$ presented us with an issue called
``proton spin crisis". Since then  many ideas have been proposed
as  solutions. One simple and plausible explanation was that there exists an
anomalous gluon  contribution to the first moment~\cite{AR}-\cite{ET}
originating from the QCD axial anomaly. This explanation was
later~\cite{BQ} supported with a notion of
the factorization-scheme dependence.
There is a set of the factorization schemes in which we obtain
\be
   \Delta B_{G}^{n=1}=-2N_f, \qquad \Delta
\gamma^{(1),n=1}_{\psi\psi }=0~.
\label{CIlike}
\ee
Let us call them CI-like schemes. In this paper we consider
four CI-like schemes, in which we take $\Delta z(n,a)=\Delta {\hat z}(n,a)$,
since both QCD and QED anomalies originate from the similar triangle diagrams.
With this choice, the relation
between the one-loop gluon and photon coefficient functions,
which holds in the $\overline {\rm MS}$ scheme, also holds in the
CI-like schemes,
\be
     \Delta B_{\gamma,~{\rm CI-like}}^n=
\frac{2}{N_f}\Delta B_{G,~{\rm CI-like}}^n~.\label{GgammaCIlike}
\ee
Thus, in addition to the relations in Eq.(\ref{CIlike}),
we obtain in the CI-like schemes
\be
\Delta B_{\gamma,~{\rm CI-like}}^{n=1}=-4~, \qquad
\quad \Delta A_{\psi,~{\rm CI-like}}^{n=1}=
\Delta A_{NS,~{\rm CI-like}}^{n=1}=0~.
\label{CIlike2}
\ee

\noindent
(i) [The chirally invariant (CI)  scheme]
\ \ In this scheme the factorization of the photon-gluon (photon-photon) cross
section into the hard and soft parts is made so that  chiral symmetry is
respected \cite{HYC,MT} and
the QCD and QED anomaly effects are absorbed into the gluon and photon 
coefficient
functions. Thus the spin-dependent quark distributions in
the CI scheme are  anomaly-free.
The transformation from the
$\overline {\rm MS}$ scheme to the CI scheme is achieved by
\be
    \Delta w(n,a={\rm CI})=0~, \qquad
    \Delta z(n,a={\rm CI})=\Delta {\hat z}(n,a={\rm
CI})=2N_f\frac{1}{n(n+1)}~.
\label{TransCI}
\ee
It has been argued by Cheng \cite{HYC} and
M\"{u}ller and Teryaev \cite{MT} that
the $x$-dependence of the axial-anomaly effect is uniquely fixed and that its
$x$-behavior  leads to the transformation rule (\ref{TransCI}) and thus to
the CI
scheme.

\bigskip
\noindent
(ii) [The Adler-Bardeen (AB)  scheme]
\ \ Ball, Forte and Ridolfi~\cite{BFR} proposed
several CI-like schemes for the analysis of the nucleon spin structure
function  $g_1(x,Q^2)$. One of them is the
Adler-Bardeen (AB) scheme which was introduced by
requiring that the change from the $\overline {\rm MS}$ scheme to this scheme
be
independent of $x$, so that the large and small $x$ behavior of the gluon
  coefficient function is unchanged. In our case, we have in moment space
\be
    \Delta w(n,a={\rm AB})=0~,\qquad
    \Delta z(n,a={\rm AB})=\Delta {\hat z}(n,a={\rm AB})=N_f\frac{1}{n}~.
\ee

\bigskip
\noindent
(iii) [The off-shell (OS)  scheme]
\ \ In this scheme~\cite{BFR} we renormalize operators while keeping the
incoming
particle
off-shell, $p^2\neq 0$, so that at renormalization (factorization) point
$\mu^2=-p^2$, the finite terms vanish.  This is exactly the same as ``the
momentum
subtraction scheme" which was used some time ago to calculate, for instance,
the polarized quark and gluon coefficient
functions~\cite{KMSU,JK}.
The CI-relations in Eqs.(\ref{CIlike}) and (\ref{CIlike2}) also hold in the OS
scheme~\cite{JKadded},  since the axial anomaly appears as a finite term in the
calculation of the triangle graph for $j_5^{\mu}$ between external gluons
(photons) and the finite term is thrown away in this scheme.
The transformation  from the
$\overline{\rm MS}$ scheme  to the OS scheme is made by choosing
\bea
    \Delta w(n,a={\rm OS})&=& C_F \biggl\{\Bigl[S_1(n)\Bigr]^2+3S_2(n)-
S_1(n)\Bigl(\frac{1}{n}- \frac{1}{(n+1)}  \Bigr)  \nonumber  \\
& & \qquad \qquad-\frac{7}{2} +\frac{2}{n}
-\frac{3}{n+1}-\frac{1}{n^2}+\frac{2}{(n+1)^2}\biggr\}~,\nonumber \\
    \Delta z(n,a={\rm OS})&=&\Delta {\hat z}(n,a={\rm OS}) \nonumber \\
&=&N_f\biggl\{
-\frac{n-1}{n(n+1)}S_1(n)+ \frac{1}{n} +\frac{1}{n^2}-\frac{4}{(n+1)^2}
\biggr\}~.
\eea
It is noted that in the OS scheme we have
$ \Delta A^n_{\psi,~{\rm OS}}=\Delta A^n_{NS,~{\rm OS}}=0 $
not only for $n=1$ but also for all $n$.

\bigskip
\noindent
(iv) [The Altarelli-Ross (AR)  scheme]
\ \
Using massive quark as a regulator for  collinear divergence,
Altarelli and Ross \cite{AR, V2} derived the same one-loop gluon coefficient
function
$\Delta C^{\gamma}_G$ as in the case of CI scheme. In order to obtain the
one-loop
quark coefficient function in this scheme, however, we need to do an extra
subtraction
so that the conservation of the nonsinglet axial currents is
secured \cite{SWV}.
The transformation rule is
\bea
    \Delta w_S(n,a={\rm AR})
&=&C_F\biggl\{2\Bigl[S_1(n)\Bigr]^2+2S_2(n)-
S_1(n)\Bigl(\frac{2}{n}- \frac{2}{n+1} +2 \Bigr) \nonumber \\
& & \qquad \qquad -2
+\frac{1}{n}-\frac{1}{n+1}+\frac{2}{(n+1)^2}\biggr\}~,  \\
    \Delta z(n,a={\rm AR})&=&\Delta {\hat z}(n,a={\rm
AR})=2N_f\frac{1}{n(n+1)}~.
\eea

\subsection{The ${\rm DIS}_{\gamma}$  scheme}

An interesting factorization scheme, which is called  ${\rm
DIS}_{\gamma}$,   was introduced some time ago into the NLO analysis of the
unpolarized real photon structure function $F_2^{\gamma}(x, Q^2)$.
Gl{\"u}ck, Reya and Vogt~\cite{GRV} observed that,  in the $\overline {\rm MS}$
scheme, the ${\rm ln}(1-x)$ term in the photonic coefficient function
$C_2^{\gamma}(x)$  for $F_2^{\gamma}$, which becomes negative and divergent for
$x \rightarrow 1$,  drives the `pointlike' part of $F_2^{\gamma}$  to large
negative values as $x \rightarrow 1$,  leading to a strong difference
between the
LO and the NLO results for
$F_{2,~{\rm pointlike}}^{\gamma}$  in the large-$x$ region.
They introduced the  ${\rm DIS}_{\gamma}$  scheme in which
the photonic coefficient function $C_2^{\gamma}$, i.e., the direct-photon
contribution to $F_2^{\gamma}$, is absorbed into the photonic quark
distributions.  It is noted that, for the real photon target,
the structure function $F_2^{\gamma}$ is decomposed into a
`pointlike' and a `hadronic' part, the former being perturbatively
calculable but not the latter.  And beyond the LO both the
`pointlike' and  the  `hadronic' parts depend on the factorization scheme
employed.
A similar situation occurs in the polarized case, and the ${\rm DIS}_{\gamma}$
scheme was applied to the NLO analysis for the
spin-dependent structure function $g_1^{\gamma}(x, Q^2)$ of the real photon
target by Stratmann and Vogelsang~\cite{SV}.

In the polarized version of ${\rm DIS}_{\gamma}$ scheme we take
\bea
\Delta w_S(n,{\rm DIS}_{\gamma})&=&\Delta w_{NS}(n,{\rm DIS}_{\gamma})=
\Delta z(n,{\rm DIS}_{\gamma})=0 ~, \label{DISwswz}\\
\Delta {\hat z}(n,{\rm DIS}_{\gamma})&=& \frac{N_f}{4}\Delta
B_{\gamma,~\overline {\rm MS}}^n
\nonumber  \\ &=&N_f\biggl\{ -\frac{n-1}{n(n+1)}S_1(n)+ \frac{3}{n}
-\frac{4}{n+1} -\frac{1}{n^2}\biggr\}~,
\eea
so that
\bea
\Delta B_{\gamma,~{\rm DIS}_{\gamma}}^n&=&\Delta B_{\gamma,~\overline {\rm
MS}}^n
-\frac{4}{N_f}\Delta {\hat z}(n,{\rm DIS}_{\gamma}) \nonumber \\
&=&0~.
\eea
Note that the relation $\grave{a}$ {\it la}
Eqs.(\ref{GgammaMS}) and (\ref{GgammaCIlike}) in the
$\overline {\rm MS}$ and CI-like factorization schemes  does not hold
anymore in
this scheme, i.e.,
\be
\Delta B_{\gamma,~{\rm DIS}_{\gamma}}^n
\neq \frac{2}{N_f}\Delta B_{G,~{\rm DIS}_{\gamma}}^n ~
(= \frac{2}{N_f}\Delta B_{G,~\overline {\rm MS}}^n)~.
\ee
For $n=1$, we have
\be
\Delta {\hat z}(n=1,{\rm DIS}_{\gamma}) =0~,
\ee
and thus, together with Eq.(\ref{DISwswz}), we observe that as far as
the first moments are concerned,
${\rm DIS}_{\gamma}$ scheme gives the same results with  $\overline {\rm
MS}$. In other words, in the ${\rm DIS}_{\gamma}$ scheme, both the QCD and
QED axial anomaly effects are  retained in the quark distributions.

\bigskip

With these preparations, we now examine the factoraization scheme dependence
of the polarized parton distributions in the virtual photon.
The two-loop anomalous dimensions of the spin-dependent operators and
one-loop photon matrix elements of the hadronic operators in the
$\overline{\rm MS}$ scheme are already known.
Corresponding quantities in a particular scheme are obtained through the
transformation rules given in Eq.(\ref{Transformation}).
Inserting these quantities  into the formulas given in Appendix A, we get
the NLO predictions for the moments of polarized  parton distributions
in a particular factorization scheme.

\subsection{Gluon distribution in the virtual photon}

Let us start with the gluon distribution.
We find that all the factorization schemes which we consider in this paper
predict the same behavior for the gluon distribution up to  NLO:
\be
    \Delta G^\gamma(n,Q^2,P^2)\vert_a =\Delta
G^\gamma(n,Q^2,P^2)\vert_{\overline {\rm MS}}~,
\ee
where $a$ means factorization schemes of CI, AB, OS, AR and ${\rm
DIS}_{\gamma}$.
This can be seen from the direct calculation or from the notion that, up to
NLO, $\Delta G^\gamma\vert_a$ satisfies the same evolution equation as
$\Delta G^\gamma\vert_{\overline {\rm MS}}$  with the same initial condition
at $Q^2=P^2$, namely,~ $\Delta G^\gamma(n,P^2,P^2)\vert_a =\Delta
G^\gamma(n,P^2,P^2)\vert_{\overline {\rm MS}}=0$.

If we consider a more general factorization scheme in which the hadronic part
of $Z^{-1}_a(n,Q^2)$ in Eq.(\ref{Zinverse}) is replaced with a
new one as follows,
\be
  \left(\matrix{1-\frac{\alpha_s}{2\pi}\Delta w_S &
-\frac{\alpha_s}{2\pi} \Delta z \cr
              0 & 1} \right) \Longrightarrow
  \left(\matrix{1-\frac{\alpha_s}{2\pi}\Delta w_S &
-\frac{\alpha_s}{2\pi} \Delta z \cr
             -\frac{\alpha_s}{2\pi}\Delta u &
1-\frac{\alpha_s}{2\pi}\Delta v} \right)~,
\ee
then, in this new factorization scheme, the predicted gluon
distribution is not the same with $\Delta
G^\gamma(n,Q^2,P^2)\vert_{\overline {\rm MS}}$ in  NLO.
However, the first moment is found to be still the same. In other words, the
first moment of the gluon distribution in the virtual photon,
$\Delta G^\gamma(n=1,Q^2,P^2)$,
is factorization-scheme independent up to NLO.
This is due to the fact that the new terms, which appear by the inclusion of
$\Delta u$ and $\Delta v$,  will
be proportional to $\Delta K_{\psi}^{0, n}$ and that $\Delta K_{\psi}^{0, 
n=1}=0$.
Also inclusion of $\Delta u$ and $\Delta v$ terms in $Z^{-1}_a$
does not modify the photon structure function $g_1^{\gamma}(x, Q^2, P^2)$
itself up to NLO,  since the gluon coefficient function starts in the order
$\alpha_s$. Moreover, the quark  distributions in the virtual photon do not
change by the inclusion of $\Delta u$ and $\Delta v$ terms.

\section{The $n=1$ moments of parton distributions}
\smallskip

The first moments of polarized parton distributions in the virtual photon
are particularly interesting since they have relevance to
the QCD and QED axial anomalies.  The explicit expressions for the moments
of $\Delta q^\gamma_S$, $\Delta G^\gamma$, and $\Delta q^\gamma_{NS}$
up to NLO are given in Appendix A. We take the $n\rightarrow 1$ limit
in these expressions. Useful $n=1$ moments of one- and two-loop
anomalous dimensions, photon matrix elements, and coefficient
functions both in the $\overline {\rm MS}$ and CI-like schemes are
enumerated in Appendix B. As far as the first moments are concerned,
${\rm DIS}_{\gamma}$ scheme gives the same results with  $\overline {\rm MS}$.
Note that we have
\be
  \lambda_+^{n=1}=0~,  \qquad \lambda_-^{n=1}=-2\beta_0~, \qquad
  \lambda^{n=1}_{NS}(=\Delta\gamma_{NS}^{0,n=1})=0~,
\ee
where $\lambda_{\pm}^{n=1}$ are eigenvalues of the one-loop hadronic
anomalous dimension matrix $\Delta\gamma_{ij}^{0,n=1}$. The zero
eigenvalues $\lambda_+^{n=1}=\lambda^{n=1}_{NS}=0$ correspond to the
conservation
of the axial-vector current at one-loop order.

\subsection{The $n=1$ moment of gluon distribution}

The expressions for the moments of gluon distribution are given in Appendix
A.2.
In these expressions the factors
\be
\frac{1}{\lambda_+^{n}}~,\quad \frac{1}{2\beta_0+\lambda_-^{n}}~,\quad
\frac{1}{2\beta_0+\lambda_-^{n}-\lambda_+^{n}}
\ee
may develop singularities at $n=1$ and so we need a little care
when we deal with them. Taking the limit
of $n$ going to 1,  we  find
\bea
{\hat L}_G^{+n}&\rightarrow& 0~,\qquad {\hat L}_G^{-n}\rightarrow
{\rm finite} ~,\nonumber\\
{\hat A}_G^{+n}&\rightarrow& {\rm finite}~,\quad {\hat B}_G^{+n}\rightarrow 0~,
\quad {\hat B}_G^{-n}\rightarrow {\rm finite}~,\\
{\hat A}_G^{-n}&\rightarrow& 72\langle e^2 \rangle N_f C_F ~.\nonumber
\eea
The terms proportional to ${\hat L}_G^{-n}$, ${\hat B}_G^{-n}$, and
${\hat A}_G^{+n}$ all vanish in the $n=1$ limit, since they are multiplied
by the following vanishing factors:
\be
\left\{ 1-\left[\frac{\alpha_s(Q^2)}{\alpha_s(P^2)}\right]^
{\lambda_{-}^n/2\beta_0+1}~\right\}, \qquad
\left\{ 1-\left[\frac{\alpha_s(Q^2)}{\alpha_s(P^2)}\right]^
{\lambda_{+}^n/2\beta_0}
\right\}~. \label{vanish}
\ee
Only exception is the term proportional to ${\hat A}_G^{-n}$.
We find for $n\rightarrow 1$
\be
{\hat A}_G^{-n}\left\{ 1-\left[\frac{\alpha_s(Q^2)}{\alpha_s(P^2)}\right]^
{\lambda_{-}^n/2\beta_0}~\right\}
  \rightarrow 72\langle e^2 \rangle N_f C_F
\frac{\alpha_s(Q^2)-\alpha_s(P^2)}{\alpha_s(Q^2)}~.
\ee
Thus we obtain
\be
\Delta G^\gamma(n=1,Q^2,P^2)=\frac{12\alpha}{\pi\beta_0}
\langle e^2 \rangle N_f~\frac{\alpha_s(Q^2)-\alpha_s(P^2)}{\alpha_s(Q^2)}~.
\label{n=1Gluon}
\ee
for the first moment of the gluon distribution in the virtual
photon. It should be emphasized that the result is factorization-scheme
independent.

\subsection{The $n=1$ moment of quark distributions}

The expressions for the moments of quark distributions are given in Appendix
A.1
and A.3. In all the factorization schemes under study, i.e., $\overline {\rm
MS}$,
${\rm DIS}_{\gamma}$, and CI-like schemes, we find  for $n\rightarrow 1$,
\bea
{\hat L}_S^{+n}&\rightarrow& 0~,\quad {\hat L}_S^{-n}\rightarrow 0~,
\quad {\hat L}_{NS}^{n}\rightarrow 0~,  \nonumber  \\
{\hat A}_S^{+n}&\rightarrow& {\rm finite}~,\quad {\hat A}_S^{-n}\rightarrow 0~,
\quad {\hat A}_{NS}^{n}\rightarrow {\rm finite}\\
{\hat B}_S^{+n}&\rightarrow& 0 ~,\quad {\hat B}_S^{-n}\rightarrow {\rm
finite}~,
\quad {\hat B}_{NS}^{n}\rightarrow 0 \nonumber
\eea
The terms proportional to ${\hat A}_S^{+n}$ and ${\hat B}_S^{-n}$ are
multiplied
by the vanishing factors in Eq.(\ref{vanish}), and the ${\hat A}_{NS}^{n}$
term
multiplied by
\be
\left\{ 1-\left[\frac{\alpha_s(Q^2)}{\alpha_s(P^2)}\right]^
{\lambda_{NS}^n/2\beta_0}~\right\},
\ee
and, therefore, the ${\hat L}_S^{+n}$, ${\hat L}_S^{-n}$,
${\hat L}_{NS}^{n}$, ${\hat A}_S^{+n}$, ${\hat A}_S^{-n}$,
${\hat A}_{NS}^{n}$,  ${\hat B}_S^{+n}$, ${\hat B}_S^{-n}$, and
${\hat B}_{NS}^{n}$ terms in Eqs.(\ref{MainSinglet}) and
(\ref{MainNonsinglet}) all vanish in the $n=1$ limit.
Then the first moments of quark distributions are given by
\bea
\Delta q_S^\gamma(n=1,Q^2,P^2)&=&\frac{\alpha}{8\pi\beta_0}{\hat C}_S^{n=1}
=\frac{\alpha}{4\pi}\Delta A^{n=1}_{\psi}~,\\
\Delta q_{NS}^\gamma(n=1,Q^2,P^2)&=&\frac{\alpha}{8\pi\beta_0}
{\hat C}_{NS}^{n=1}=\frac{\alpha}{4\pi}\Delta A^{n=1}_{NS}~.
\eea
We now see that scheme dependence for the first moments
of quark distributions is coming from the photon matrix elements
$\Delta A^n_{\psi}$ and $\Delta A^n_{NS}$.

\bigskip

In the case of CI-like factorization schemes,
$a={\rm CI,AB,OS,AR}$,
we have
\bea
    \Delta w(n=1,a)&=&0,  \nonumber  \\
\Delta z(n=1,a)&=&\Delta {\hat z}(n=1,a)=N_f   \qquad {\rm for}\ a={\rm
CI,AB,OS,AR}
\eea
We find from Eqs.(\ref{TransForMat}) and (\ref{PhotonMSn=1}) that these schemes
give
\be
\Delta A_{\psi,~a}^{n=1}=\Delta A_{NS,~a}^{n=1}=0~.
\ee
This leads to an interesting result: The first moment
of spin-dependent quark distributions  in the virtual photon vanish
in  NLO for $a={\rm CI,AB,OS,AR}$.
\be
\Delta q_S^\gamma(n=1,Q^2,P^2)\vert_a =\Delta
q_{NS}^\gamma(n=1,Q^2,P^2)\vert_a
=0   \label{QSCI}
\ee
The vanishing first moments imply that the axial anomaly effects
do not reside in the quark distributions.
In these CI-like schemes, the  QCD and QED axial anomalies  are
transfered to the gluon and photon coefficient functions,
respectively, and their
first moments do not vanish. Indeed we obtain from Eqs.(\ref{aCoeff})
and (\ref{Zinverse})
\bea
\Delta C_{G,~a}^{\gamma,~n=1}&=&-\langle e^2
\rangle\frac{\alpha_s(Q^2)}{2\pi}N_f
\\
\Delta C_{\gamma,~a}^{\gamma,~n=1}&=&-\frac{3\alpha}{\pi}\langle e^4 \rangle
N_f
\left(1-\frac{\alpha_s(Q^2)}{\pi}\right)\qquad {\rm for}\ a={\rm CI,AB,OS,AR}
\label{CoeffCIlike}
\eea
where we have used the fact~\cite{KMMSU,ZN,SU}
\bea
\frac{1}{<e^2>}\Delta C_{S}^{\gamma,~n=1}\vert_{\overline{\rm MS}}&=&
    \Delta C_{NS}^{\gamma,~n=1}\vert_{\overline {\rm MS}}=
1-\frac{\alpha_s}{\pi}+ {\cal O}(\alpha^2_s) \label{n=1QuarkCoeff}\\
\Delta C_{\gamma}^{\gamma,~n=1}\vert_{\overline {\rm MS}}&=&
0+ {\cal O}(\alpha\alpha^2_s)
\eea

On the other hand, in the $\overline {\rm MS}$ (and also in ${\rm
DIS}_{\gamma}$)
we  obtain from Eq.(\ref{PhotonMSn})
\bea
\Delta A_{\psi,~\overline {\rm MS}}^{n=1}&=&-12\langle e^2 \rangle N_f  \\
\Delta A_{NS,~\overline {\rm MS}}^{n=1}&=&
-12(\langle e^4 \rangle -\langle e^2 \rangle^2) N_f
\eea
and thus $\Delta q_S^{\gamma,~n=1}\vert_{\overline {\rm MS}}$ and
$\Delta q_{NS}^{\gamma,~n=1}\vert_{\overline {\rm MS}}$
are non-zero constant. Actually we can go one step further to the order
of $\alpha_s$ QCD  corrections. This is due to the fact that, in the
$\overline {\rm MS}$ scheme, the parton distribution
$\Delta \qv^{\gamma}(n=1)\vert_{\overline {\rm MS}}=(\Delta q_S^{\gamma},~
\Delta G^{\gamma},~ \Delta q_{NS}^{\gamma})\vert_{\overline {\rm MS}}$
satisfies a  homogeneous differential equation without inhomogeneous LO
and NLO $\Delta K$ terms.  Indeed we find
\bea
   \Delta q_S^{\gamma}(n=1,Q^2,P^2)\vert_{\overline {\rm MS}}&=&
\Bigl[-\frac{\alpha}{\pi}~3 \langle e^2 \rangle N_f  \Bigr]
\left\{1-\frac{2}{\beta_0}\frac{\alpha_s(P^2)-\alpha_s(Q^2)}{\pi}
  N_f\right\}  \label{QSMS}\\
\Delta q_{NS}^{\gamma}(n=1,Q^2,P^2)\vert_{\overline {\rm MS}}&=&
\Bigl[-\frac{\alpha}{\pi}~3 \Bigl( \langle e^4 \rangle -\langle e^2 \rangle^2
\Bigr) N_f  \Bigr]
\left\{1+{\cal O}(\alpha_s^2) \right\}~,   \label{QNSMS}
\eea
the derivation of which is shown in Appendix C.
In the $\overline {\rm MS}$ scheme, the axial anomaly effects are retained
in the
quark distributions. The factors
$\Bigl[-\frac{\alpha}{\pi}~3 \langle e^2 \rangle N_f  \Bigr]$ and
$\Bigl[-\frac{\alpha}{\pi}~3 \Bigl( \langle e^4 \rangle -\langle e^2 \rangle^2
\Bigr) N_f  \Bigr]$
are related to the QED axial anomaly and a term
$\frac{2}{\beta_0}\frac{\alpha_s(P^2)-\alpha_s(Q^2)}{\pi} N_f$
in $\Delta q_S^{\gamma,~n=1}\vert_{\overline {\rm MS}}$ is coming from the
QCD axial anomaly.

\subsection{The $n=1$ moment of $ g_1^{\gamma}(x,Q^2,P^2)$}

The polarized structure function $ g_1^{\gamma}(x,Q^2,P^2)$ of the virtual
photon
satisfies the following sum rule\cite{NSV,SU}:
\bea
\int_0^1dx g_1^\gamma(x,Q^2,P^2)
&=&-\frac{3\alpha}{\pi}\langle e^4 \rangle N_f
\left(1-\frac{\alpha_s(Q^2)}{\pi}\right)       \nonumber\\
&+&\frac{6\alpha}{\pi\beta_0}\Bigl[\langle e^2 \rangle N_f \Bigr]^2~
\frac{\alpha_s(P^2)-\alpha_s(Q^2)}{\pi}
+{\cal O}(\alpha_s^2). \label{Oalpha}
\eea
This sum rule is of course the factorization-scheme independent. Now we
examine
how the scheme-dependent parton distributions contribute to this sum rule.
In the
CI-like schemes ($a={\rm CI,AB,OS, AR}$), the first moment of the quark
distributions
vanish in  NLO, and thus the contribution to the sum rule comes from the
gluon and
photon  distributions. Equations (\ref{n=1Gluon}) and
(\ref{CoeffCIlike}) show that
\be
\Delta C_{G,~a}^{\gamma,~n=1}\Delta G^\gamma(n=1,Q^2,P^2)\vert_{a}+
\Delta C_{\gamma,~a}^{\gamma,~n=1}
\ee
leads to the result (\ref{Oalpha}).
On the other hand, in the $\overline {\rm MS}$ scheme
(and also in ${\rm DIS}_{\gamma}$),
the one-loop gluon and
photon
coefficient functions vanish, $\Delta B_{G,~\overline {\rm MS}}^{n=1}=\Delta
B_{\gamma,~\overline {\rm MS}}^{n=1}=0$ and, therefore, the sum rule is derived
from the
quark contributions. Indeed  we find from Eqs.(\ref{n=1QuarkCoeff}),
(\ref{QSMS}-\ref{QNSMS})
\be
\Delta C_{S,~\overline {\rm MS}}^{\gamma,~ n=1}~
  \Delta q_S^{\gamma}(n=1,Q^2,P^2)\vert_{\overline {\rm MS}}~+
\Delta C_{NS,~\overline {\rm MS}}^{\gamma,~ n=1}~
\Delta q_{NS}^{\gamma}(n=1,Q^2,P^2)\vert_{\overline {\rm MS}}
\ee
leads to the same result.

It is interesting to note that the sum rule (\ref{Oalpha}) is the
consequence of the QCD and QED
axial anomalies and that in the CI-like schemes the anomaly effect resides
in the gluon contribution while, in  $\overline {\rm MS}$, in the quark
contributions.
Furthermore, the first term of the sum rule (\ref{Oalpha}) is coming from
the QED
axial anomaly and the second is from the QCD axial anomaly\footnote{This notion
was first pointed out by the authors of Ref.\cite{NSV}}.

\section{Behaviors of parton distributions near $x=1$}
\smallskip

The behaviors of parton distributions near~$x=1$
are governed by the large-$n$ limit of those moments.
In the leading order,
parton distributions are factorization-scheme independent. For large $n$,~
$\Delta q_S^{\gamma}(n,Q^2,P^2)\vert_{\rm LO}$ and $\Delta
q_{NS}^{\gamma}(n,Q^2,P^2)\vert_{\rm LO}$ behave as $1/(n~{\rm ln}~n)$, while
$\Delta G^{\gamma}(n,Q^2,P^2)\vert_{\rm LO}\propto 1/(n~{\rm ln}~n)^2$.
Thus in $x$ space,  the parton distributions vanish for $x \rightarrow 1$.
In fact, we find
\bea
  \Delta q_S^{\gamma}(x,Q^2,P^2)\vert_{\rm LO}&\approx&
\frac{\alpha}{4\pi}\frac{4\pi}{\alpha_s(Q^2)}
N_f\langle e^2 \rangle\frac{9}{4}~\frac{-1}{{\rm ln}~(1-x)}~, \label{QuarkLO}\\
\Delta G^{\gamma}(x,Q^2,P^2)\vert_{\rm LO}&\approx&
\frac{\alpha}{4\pi}\frac{4\pi}{\alpha_s(Q^2)}
N_f\langle e^2 \rangle\frac{1}{2}~\frac{-{\rm ln}~x}{{\rm ln}^2~(1-x)}~.
\eea
The behaviors of $\Delta q_{NS}^{\gamma}(x,Q^2,P^2)$ for $x \rightarrow 1$,
both in LO and NLO, are always
given by the corresponding expressions for $\Delta q_{S}^{\gamma}(x,Q^2,P^2)$
with  replacement of the charge factor $\langle e^2 \rangle$ with
$(\langle e^4 \rangle -\langle e^2 \rangle^2)$.

In the $\overline {\rm MS}$ scheme,
the moments of the NLO parton distributions are written in large $n$ limit as
\bea
\Delta q_S^{\gamma}(n,Q^2,P^2)\vert_{{\rm NLO},~\overline {\rm MS}}
&\longrightarrow& \frac{\alpha}{4\pi} N_f\langle e^2 \rangle 6~\frac{{\rm
ln}~n}{n}~,      \\
\Delta G^{\gamma}(n,Q^2,P^2)\vert_{{\rm NLO},~\overline {\rm MS}}
&\longrightarrow& \frac{\alpha}{4\pi} N_f\langle e^2 \rangle 3~\frac{1}{n^2}~.
\eea
So we have near $x=1$
\bea
    \Delta q_S^{\gamma}(x,Q^2,P^2)\vert_{{\rm NLO},~\overline {\rm MS}}&\approx&
\frac{\alpha}{4\pi} N_f\langle e^2 \rangle 6~\Bigl[-{\rm
ln}(1-x)\Bigr]~,\label{NLOMSbar}\\
\Delta G^{\gamma}(x,Q^2,P^2)\vert_{{\rm NLO},~\overline {\rm MS}}&\approx&
\frac{\alpha}{4\pi} N_f\langle e^2 \rangle 3~\Bigl[-{\rm ln}~x\Bigr] ~.
\eea
It is remarkable that, in the $\overline {\rm MS}$ scheme, quark parton
distributions,
$\Delta q_S^{\gamma}(x)\vert_{{\rm NLO},~\overline {\rm MS}}$ and
$\Delta q_{NS}^{\gamma}(x)\vert_{{\rm NLO},~\overline {\rm MS}}$
positively diverge as $[-{\rm ln}(1-x)]$ for $x\rightarrow 1$.
Recall that $\Delta G^{\gamma}(x,Q^2,P^2)\vert_{\rm NLO}$ is the same among
the schemes which we consider in this paper.
The NLO quark distributions in the CI, AB, AR and ${\rm DIS}_{\gamma}$ schemes
also diverge as $x\rightarrow 1$, since their moments behave
as ${\rm ln}~n/n$ in the large $n$-limit.
We find for large $x$,
\bea
\Delta q_S^{\gamma}(x,Q^2,P^2)\vert_{{\rm NLO},~{\rm CI}}&\approx&
\frac{\alpha}{4\pi} N_f\langle e^2 \rangle 6~\Bigl[-{\rm
ln}(1-x)\Bigr]~,   \label{NLOCI}\\
\Delta q_S^{\gamma}(x,Q^2,P^2)\vert_{{\rm NLO},~{\rm AB}}&\approx&
\frac{\alpha}{4\pi} N_f\langle e^2 \rangle 6~\Bigl[-{\rm ln}(1-x) +2\Bigr]~,
\label{NLOAB}\\
\Delta q_S^{\gamma}(x,Q^2,P^2)\vert_{{\rm NLO},~{\rm AR}}&\approx&
\frac{\alpha}{4\pi} N_f\langle e^2 \rangle 18~\Bigl[-{\rm ln}(1-x)\Bigr]~,
\label{NLOAR} \\
\Delta q_S^{\gamma}(x,Q^2,P^2)\vert_{{\rm NLO},~{\rm DIS}_{\gamma}}&\approx&
\frac{\alpha}{4\pi} N_f\langle e^2 \rangle 6~{\rm ln}(1-x)~.
\label{NLODISgamma}
\eea
It is noted that $\Delta q_S^{\gamma}(x,Q^2,P^2)\vert_{{\rm NLO},~{\rm
DIS}_{\gamma}}$ negatively diverges as $x\rightarrow 1$. This is due to the
fact that the photonic coefficient function $\Delta C_\gamma^{\gamma}(x)$,
which in $\overline {\rm MS}$ becomes negative and divergent for
$x\rightarrow 1$ , is absorbed into the quark distributions in the ${\rm
DIS}_{\gamma}$ scheme.

On the other hand, the OS scheme gives quite different behaviors near $x=1$
for
the quark distributions. Since the typical two-loop anomalous dimensions in
the
OS scheme behave in the large $n$-limit as
\be
\Delta  \gamma^{(1),n}_{NS,~{\rm OS}}\sim \Delta
\gamma^{(1),n}_{qq,~{\rm OS}}\propto
{\rm ln}^2~ n~, \qquad \Delta K^{(1),n}_{S,~{\rm OS}} \propto\frac{{\rm
ln}~n}{n}~,
\ee
while in the  $\overline {\rm MS}$ scheme
\be
\Delta  \gamma^{(1),n}_{NS,~\overline {\rm MS}}\sim \Delta
\gamma^{(1),n}_{qq,~\overline {\rm MS}}\propto {\rm ln}~ n~, \qquad
\Delta K^{(1),n}_{S,~\overline {\rm MS}} \propto\frac{{\rm ln}^2~n}{n}~,
\ee
we find that the moment of $\Delta q_{S}^\gamma(n,Q^2,P^2)\vert_{\rm NLO}$ in
the OS
scheme is expressed  in the large $n$-limit as
\be
\Delta q_{S}^\gamma(n,Q^2,P^2)\vert_{{\rm NLO},~{\rm OS}}
\longrightarrow  \frac{\alpha}{4\pi}N_f \langle e^2 \rangle
\Bigl[\frac{69}{8} + \frac{3}{4}N_f \Bigr]\frac{1}{n}
\ee
In $x$ space,
$\Delta q_{S}^\gamma(x,Q^2,P^2)\vert_{{\rm NLO},~{\rm OS}}$
does not diverge for $x \rightarrow 1$ but approaches a constant value:
\be
\Delta q_{S}^\gamma(x,Q^2,P^2)\vert_{{\rm NLO},~{\rm OS}}
\longrightarrow \frac{\alpha}{4\pi}N_f \langle e^2 \rangle
\Bigl[\frac{69}{8} + \frac{3}{4}N_f \Bigr]~. \label{NLOOS}
\ee
Therefore, as far as the
large $x$-behaviors of quark distributions, and gluon and photon
coefficient functions (see Eqs.(\ref{xCoeffiGluon}-
\ref{xCoeffiGamma}) below) are concerned, the OS
scheme is more appropriate than other
schemes in the sense that they remain finite.
Also the quark coefficient function in the OS scheme
has a milder divergence for $x\rightarrow 1$ than those predicted
in other schemes (see Eq.(\ref{xCoeffiQuark})).

Before ending this section, we now
show that, as $x \rightarrow 1$, the polarized virtual
photon structure function $g_1^{\gamma}(x, Q^2, P^2)$ approaches a constant
value
\be
   \kappa=\frac{\alpha}{4\pi}N_f\langle e^4 \rangle
\Bigl[-\frac{51}{8}+\frac{3}{4}N_f \Bigr]~,
\label{kappa}
\ee
in NLO.  The result is, of course, factorization-scheme independent.
It is interesting to note that the constant value $\kappa$ coincides exactly
with the one given in Eq.(4.39) of Ref.\cite{KS2},
which was derived as the large $n$ limit of the moment of
the NLO term $b_2(x)$ for the unpolarized structure
function $F_2^{\gamma}$~\cite{BB}.
In the leading order, Eq.(\ref{QuarkLO}) tells us that
\bea
  g_1^{\gamma}(x,Q^2,P^2)\vert_{\rm LO}&=&
\langle e^2 \rangle\Delta q_S^{\gamma}(x,Q^2,P^2)\vert_{\rm LO}+
\Delta q_{NS}^{\gamma}(x,Q^2,P^2)\vert_{\rm LO}  \nonumber \\
&\longrightarrow &
\frac{\alpha}{4\pi}\frac{4\pi}{\alpha_s(Q^2)}
N_f\langle e^4 \rangle \frac{9}{4}~\frac{-1}{{\rm ln}~(1-x)}~,
\eea
and thus $ g_1^{\gamma}(x,Q^2,P^2)\vert_{\rm LO}$ vanishes as $x \rightarrow
1$.

In order to analyze the large $x$-bahavior of the next-leading order
$g_1^{\gamma}(x,Q^2,P^2)\vert_{\rm NLO}$,
we need information on the coefficient
functions. Note that $\Delta B_{NS}^{n}\vert_a=\Delta B_S^{n}\vert_a$.
They behave, as $x \rightarrow 1$,
\bea
\Delta B_S(x)\vert_a &\longrightarrow&
\cases{2C_F~\biggl[\frac{2~{\rm ln}(1-x)}{1-x}\biggr]_+
  & for\ \  $a=\overline {\rm MS},~{\rm CI,~ AB},~{\rm DIS}_{\gamma}$~, \cr
\cr
  -2C_F~\biggl[\frac{2~{\rm ln}(1-x)}{1-x}\biggr]_+
& for\ \  $a={\rm AR}$~,  \cr
\cr
3C_F~\frac{-1}{(1-x)_+}& for\ \  $a={\rm OS}$~,   } \label{xCoeffiQuark}\\
&& \nonumber  \\
\Delta B_G(x)\vert_a &\longrightarrow&
\cases{2N_f~{\rm ln}(1-x) & for\ \  $a=\overline {\rm MS},~
{\rm CI,~ AB,~ AR},~{\rm DIS}_{\gamma}$~, \cr
\label{xCoeffiGluon} \cr
  -4N_f& for\ \  $a={\rm OS}$~, } \\
&& \nonumber  \\
\Delta C(x)_{\gamma}^{\gamma}\vert_a &\longrightarrow&
\cases{\frac{\alpha}{4\pi}\langle e^4 \rangle 12N_f~{\rm ln}(1-x)
& for\ \  $a=\overline {\rm MS},~  {\rm CI,~ AB,~ AR}$~, \cr
    \label{xCoeffiGamma}\cr
  -\frac{\alpha}{4\pi}\langle e^4 \rangle 24N_f~& for\ \  $a={\rm OS}$~, \cr
\cr
  0 & for\ \ $a={\rm DIS}_{\gamma}$~. }
\eea
The coefficient functions $\Delta B_S(x)$ and $\Delta B_G(x)$ are the
same that appear in the polarized nucleon structure function
$g_1(x,Q^2)$. The $\Delta B_S(x)$ in all schemes considered here
diverges as $x \rightarrow 1$, but
OS scheme gives a milder divergence for $\Delta B_S(x)$
than other schemes.
Also note that $\Delta B_G(x)\vert_{OS}$ remains finite as
$x \rightarrow 1$, but $\Delta B_G(x)$ in other schemes negatively
diverge.

  Let us write $g_1^{\gamma}(x,Q^2,P^2)\vert_{\rm NLO}$ in
terms of partonic contributions as follows:
\be
g_1^{\gamma}(x,Q^2,P^2)\vert_{\rm NLO}=
g_1^{\gamma}(x)\vert_{\rm NLO}^{\rm quark}
+ g_1^{\gamma}(x)\vert_{\rm NLO}^{\rm gluon}+
\Delta C_{\gamma}^{\gamma}(x)~,
\ee
where
\bea
g_1^{\gamma}(x)\vert_{\rm NLO}^{\rm quark}&\equiv&
\langle e^2 \rangle\Delta q_S^{\gamma}(x,Q^2,P^2)\vert_{\rm NLO}
+\Delta q_{NS}^{\gamma}(x,Q^2,P^2)\vert_{\rm NLO}\nonumber \\
& &+\langle e^2 \rangle\frac{\alpha_s(Q^2)}{4\pi}\Delta B_S(x)~\otimes
\Delta q_S^{\gamma}(x,Q^2,P^2)\vert_{\rm LO}\nonumber \\
& &+\frac{\alpha_s(Q^2)}{4\pi}\Delta B_{NS}(x)~\otimes
\Delta q_{NS}^{\gamma}(x,Q^2,P^2)\vert_{\rm LO} ~, \label{g1QuarkContri}  \\
g_1^{\gamma}(x)\vert_{\rm NLO}^{\rm gluon}&\equiv&
\langle e^2 \rangle\frac{\alpha_s(Q^2)}{4\pi}\Delta B_G(x)\otimes
\Delta G^{\gamma}(x,Q^2,P^2)\vert_{\rm LO}~.
\eea
Then we find, for $x \rightarrow 1$,
\bea
g_1^{\gamma}(x)\vert_{\rm NLO}^{\rm quark} &\longrightarrow&
\cases{-\frac{\alpha}{4\pi}\langle e^4 \rangle 12N_f\ln(1-x)
  & for\ \  $a=\overline
{\rm MS},~  {\rm CI,~ AB,~AR}$ ~,\cr
\label{QuarkContriNLO} \cr
\frac{\alpha}{4\pi}\langle e^4 \rangle N_f\Bigl[ \frac{141}{8} +\frac{3}{4} N_f
\Bigr]& for\ \  $a={\rm OS}$~, \cr
\cr
\frac{\alpha}{4\pi}\langle e^4 \rangle N_f\Bigl[ -\frac{51}{8} +\frac{3}{4} N_f
\Bigr]& for\ \  $a={\rm DIS}_{\gamma}$~, }~
\eea
The NLO gluon contribution
$g_1^{\gamma}(x,Q^2,P^2)\vert_{\rm NLO}^{\rm gluon}$ vanishes faster than
$({\rm ln}~x)^2$ in any scheme under consideration. As for the NLO quark
contribution, $g_1^{\gamma}(x,Q^2,P^2)\vert_{\rm NLO}^{\rm quark}$ in
$\overline{\rm MS}$, CI, AB, AR schemes,  diverges as
$[-{\rm ln}(1-x)]$ for $x\rightarrow 1$. However, Eq.(\ref{xCoeffiGamma})
shows that the one-loop photon coefficient
function $\Delta C^{\gamma}_{\gamma}(x)$ in these schemes also diverges as
$[-{\rm ln}(1-x)]$ with the opposite sign and the sum becomes finite.
On the other hand, in the OS scheme, we observe from
Eqs.(\ref{QuarkContriNLO}) and (\ref{xCoeffiGamma}) that
both the quark contribution  and photon
coefficient function remain  finite as $x \rightarrow 1$, and
it is easily seen that the sum
\be
g_1^{\gamma}(x)\vert_{\rm NLO,~ OS}^{\rm quark}+
\Delta C(x)_{\gamma,~{\rm OS}}^{\gamma}
\ee
approaches the constant value
$\kappa$ given in Eq.(\ref{kappa}).
In the ${\rm DIS}_{\gamma}$ scheme, the  NLO quark
contribution $g_1^{\gamma}(x)\vert_{{\rm
NLO},~{\rm DIS}_{\gamma}}^{\rm quark}$ reaches the finite value $\kappa$
as  $x\rightarrow 1$,
since $\Delta C(x)_{\gamma,~ {\rm DIS}_{\gamma}}^{\gamma}\equiv 0$.
In fact, as we see from Eq.(\ref{g1QuarkContri}),
$g_1^{\gamma}(x)\vert_{{\rm NLO}}^{\rm quark}$ is  made up of two
parts,
the one from $\Delta q_S^{\gamma}(x,Q^2,P^2)\vert_{{\rm NLO}}$ and the other
from $\Delta B_S(x)\otimes
\Delta q_S^{\gamma}(x,Q^2,P^2)\vert_{{\rm LO}} $,  plus their
non-singlet quark counterparts. In ${\rm DIS}_{\gamma}$, both contributions
diverge as $x\rightarrow 1$, but with the opposite sign,  and the sum
remains finite.

The constant value $\kappa$  in Eq.(\ref{kappa}) is
negative unless
$N_f\geq 9$. Consequently, it seems superficially that QCD with 8 flavors or
less
predicts that the structure  function $g_1^{\gamma}(x,Q^2,P^2)$ turns out to be
negative for $x$ very close to 1,  since the leading term
$g_1^{\gamma}(x,Q^2,P^2)\vert_{\rm LO}$ vanishes as $x \rightarrow 1$.
But the fact is that $x$ cannot reach exactly one. The constraint
$(p+q)^2\geq 0$ gives
\be
x\leq x_{\rm max}=\frac{Q^2}{Q^2+P^2}~,
\ee
and we find
\be
g_1^{\gamma}(x=x_{\rm max},Q^2,P^2)\vert_{\rm LO}
\quad >\quad \frac{\alpha}{4\pi}N_f\langle e^4 \rangle \frac{3}{C_F}\beta_0
\ee
and the sum $g_1^{\gamma}(x=x_{\rm max},Q^2,P^2)\vert_{\rm LO+NLO}$
is indeed positive.

\section{Numerical analysis}
\smallskip

The parton distribution functions are recovered from the moments by the
inverse Mellin
transformation. In Fig. 2 we plot the factorization scheme dependence of
the singlet quark
distribution  $\Delta q_S^{\gamma}(x,Q^2,P^2)$ beyond the LO in units of \\
$(3N_f \langle e^2 \rangle\alpha/\pi){\rm ln}(Q^2/P^2)$. We have taken $N_f=3$,
$Q^2=30~ {\rm GeV}^2$, $P^2=1~ {\rm GeV}^2$,  and the QCD scale parameter
$\Lambda=0.2~ {\rm GeV}$. All four
CI-like (i.e., CI, AB, OS and AR) curves cross the $x$-axis nearly at the same
point, just below
$x=0.5$, while the $\overline {\rm MS}$ curve crosses at above $x=0.5$. This is
understandable
since we saw from Eqs.(\ref{QSCI}, \ref{QSMS}) that the first moment of
$\Delta q_S^\gamma$ vanishes in the CI-like schemes while it is negative in the
$\overline {\rm MS}$ scheme. The ${\rm DIS}_{\gamma}$ curve crosses the
$x$-axis
below $x=0.5$, though the first moment of
$\Delta q_S^\gamma\vert_{{\rm DIS}_{\gamma}}$ is negative, taking the same
value
with the one in the $\overline {\rm MS}$ scheme. Comparing the ${\rm
DIS}_{\gamma}$ curve at large $x$ with the $\overline {\rm MS}$ one, we will
see that rapid dropping of the ${\rm DIS}_{\gamma}$ curve  as $x\rightarrow
1$ drives the crossing point  below $x=0.5$.

As $x\rightarrow 1$, we observe that the $\overline{\rm MS}$, CI, AB, and AR
curves  continue to increase. In fact we see that the $\overline {\rm MS}$
and CI
curves tend to merge, the AB curve comes above those two curves
and the AR curve diverges more rapidly than the other three.
On the other hand, the OS and ${\rm DIS}_{\gamma}$ curves start to drop
at large $x$. The OS curve continues to increase till near $x=1$, and then
starts to drop to reach a  finite positive value. The ${\rm DIS}_{\gamma}$
curve
reaches maximum  at $x\approx 0.8$ and drops to negative values.
These behaviors are inferred from
Eqs.(\ref{NLOMSbar}, \ref{NLOCI}-\ref{NLODISgamma}, \ref{NLOOS}).

Concerning the non-singlet quark distribution $\Delta
q_{NS}^{\gamma}(x,Q^2,P^2)$,
we find that when we take  into account the charge factors, it falls on the
singlet quark distribution  in almost all $x$ region; namely two ``normalized"
distributions $\Delta {\widetilde q}_S^{\gamma}\equiv
\Delta q_S^{\gamma}/\langle e^2 \rangle$ and
$\Delta{\widetilde q}_{NS}^{\gamma}\equiv \Delta
q_{NS}^{\gamma}/(\langle e^4 \rangle -\langle e^2 \rangle^2)$  mostly overlap
except at very small $x$ region. The situation is the same in all
factorization schemes we have studied in this paper. This is attributable to
the fact that once the charge factors are taken into account, the evolution
equations for both  $\Delta {\widetilde q}_S^{\gamma}$
and $\Delta{\widetilde q}_{NS}^{\gamma}$ have the same inhomogeneous LO and
NLO $\Delta K$ terms and the same initial conditions at $Q^2=P^2$
(see Eq.(\ref{ChargeFactor})).

In Fig. 3 we plot again the OS and ${\rm DIS}_{\gamma}$ predictions for
$\Delta q_S^{\gamma}(x,Q^2,P^2)$  together with the LO result.
The motivation of having introduced ${\rm DIS}_{\gamma}$ scheme
into the analysis of the unpolaorized (polarized) real photon
structure function $F_2^{\gamma}$ ($g_1^{\gamma}$) was to reduce
the discrepancies at large-$x$ region between the LO and the NLO results for
the `pointlike' part of $F_2^{\gamma}$ ($g_1^{\gamma}$). When applied to the
polarized virtual photon case, it is seen from Fig. 2 and 3 that
${\rm DIS}_{\gamma}$ scheme gives a better behavior for
$\Delta q_S^{\gamma}(x,Q^2,P^2)$ at large $x$  than   $\overline {\rm MS}$
in the sense that ${\rm DIS}_{\gamma}$ curve is closer to the LO result.
However, we observe that absorbing the photonic coefficient function
$\Delta C^{\gamma}_{\gamma}$ into the quark distributions in the
${\rm DIS}_{\gamma}$ scheme has too much effect on their large-$x$ behaviors:
The ${\rm DIS}_{\gamma}$ curve for
$\Delta q_S^{\gamma}(x,Q^2,P^2)$ goes under the LO one at $x\approx 0.6$ and
the difference between the two grows as $x\rightarrow 1$.
In fact the ${\rm DIS}_{\gamma}$ curve drops to negative values near
at $x=1$.

\relax From the viewpoint of `perturbative stabilities' we find that
the OS curve shows more appropriate behavior than the others.
We see from Fig. 3 that the differences between the OS and LO curves
are very small for the range $0.05<x<0.7$. And the OS curve
comes above the LO for $x>0.7$.

Fig. 4  shows the $Q^2$-dependence of $\Delta q_S^{\gamma}(x,Q^2,P^2)$ in
the OS scheme in units of $(3 N_f \langle e^2 \rangle\alpha /\pi){\rm
ln}(Q^2/P^2)$.
Three curves with $Q^2=30,~ 50$ and
$100~ {\rm GeV}^2$ almost overlap in whole $x$ region except in the vicinity
of $x=1$. We see from Fig. 4 that, in the OS scheme, $\Delta q_S^{\gamma}$
beyond the LO   behaves approximately as the one obtained from the box (tree)
diagram calculation,
\be
   \Delta q_S^{\gamma({\rm Box})}(x,Q^2,P^2)=(2x-1) 3 N_f
\langle e^2 \rangle\frac{\alpha}{\pi}~{\rm ln}~
\frac{Q^2}{P^2}~.
\ee

The gluon distribution $\Delta G^{\gamma}(x,Q^2,P^2)$
beyond the LO is shown in Fig. 5 in units of $(3 N_f \langle e^2
\rangle\alpha /\pi){\rm ln}(Q^2/P^2)$, with three different $Q^2$
values. Recall that every scheme considered in this paper
predicts the same behavior for the gluon distribution up to NLO.
We do not see much difference in three curves with different $Q^2$.
This means the $\Delta G^{\gamma}$
is approximately proportional to ${\rm ln}(Q^2/P^2)$~.
But, compared with quark distributions, $\Delta G^{\gamma}$
is very much small in absolute value  except at the small $x$ region.

In Fig. 6  we plot the virtual photon structure function
$g_1^\gamma(x,Q^2,P^2)$ in the NLO
for $N_f=3$, $Q^2=30$ GeV$^2$ and $P^2=1$ GeV$^2$ and the QCD scale parameter
$\Lambda=0.2$ GeV. The vertical axis corresponds to
\be
g_1^\gamma(x,Q^2,P^2)/\frac{3\alpha}{\pi}N_f<e^4>\ln\frac{Q^2}{P^2}.
\label{normalized}
\ee
Also shown are  the LO result,  the Box (tree) diagram contribution,
\be
g_1^{\gamma({\rm Box})}(x,Q^2,P^2)
=(2x-1)\frac{3\alpha}{\pi}N_f<e^4>\ln\frac{Q^2}{P^2}~,
\ee
and the Box diagram contribution including non-leading (NL) correction
with mass being ignored
\be
g_1^{\gamma({\rm Box(NL)})}(x,Q^2,P^2)=\frac{3\alpha}{\pi}N_f<e^4>
\left[(2x-1)\ln\frac{Q^2}{P^2}-2(2x-1)(\ln x+1)\right]~.
\ee
In our previous paper \cite{SU}, there was an error in the
program for numerical evaluation of the NLO $g_1^\gamma(x,Q^2,P^2)$. The
corrected graph (NLO curve) here is different from the
corresponding one  in Fig.2 of Ref.\cite{SU}. The new NLO curve
appears lower than the previous one for $x<0.7$ and rather enhanced above
$x=0.7$.  We observe that the corrected NLO curve remains below the LO one,
and that the NLO QCD corrections are significant at large
$x$ as well as at low $x$.

For the case of the real photon target, $P^2=0$, the structure function can be
decomposed as
\be
g_1^\gamma(x,Q^2)=g_1^\gamma(x,Q^2)\vert_{\rm pert.} +
g_1^\gamma(x,Q^2)\vert_{\rm non-pert.}~.
\ee
The first term, the point-like piece, can be calculated in a perturbative
method. Actually, it can be obtained by setting $P^2=\Lambda^2$
in the expressions of parton distributions in
Eq.(\ref{Solg}) or (\ref{gonegamma}).
The second term can only be computed by some non-perturbative methods.
In Fig.7, we plot the point-like piece of the real
photon  $g_1^\gamma(x,Q^2)$ in the NLO, together with the LO result
and the Box (tree) diagram contribution.
The NLO curve, which is calculated by the corrected computer program, is
different from the previous one in Fig. 6 in ref.\cite{SU}. The new NLO curve
appears lower than the previous one for $x<0.6$ and  enhanced above
$x=0.6$. Also it remains below the LO curve.
The NLO result qualitatively consistent with
the analysis by Stratmann and Vogelsang \cite{SV}.
In the unpolarized case, the moment of $F_2^\gamma$ has a singularity
at $n=2$ which leads to the negative structure function at low $x$.
Thus we need
some regularization prescription to recover positive structure function
as discussed in Refs. \cite{BILL,AG,DUKE}.
Note that we do not have such complication at $n=1$ for the polarized
case.

Finally, in our numerical analyis, we took $P^2=1 {\rm GeV}^2$, which
may not be necessarily large enough for the non-perturbative effects to
be dying away. For our {\it normalized} parton distributions, however, the 
larger
values of $P^2$ would not give any sizable change in shape and magnitude.

\section{Conclusion}
\smallskip

In the present paper, we have studied in detail the spin-dependent
parton distributions inside the virtual photon, which
can be predicted entirely up to NLO in  perturbative QCD.
The virtual photon target provides a good testing ground for examining
the factorization scheme dependence of the quark and gluon distributions.
We have investigated the polarized parton distributions in several
different factorization schemes. We derived the explicit transformation
rules from one scheme to another for the
coefficient functions, the finite photon matrix elements and
the two-loop anomalous dimensions or parton splitting functions.

In particular, we studied the QCD and QED axial anomaly effects on the 
first moments
of quark distributions
to see the interplay between the axial anomalies and factorization
schemes. We find that, in the CI-like schemes, the first moments of polarized
quark distributions, both flavor singlet and non-singlet, vanish in NLO while
the standard $\overline {\rm MS}$ scheme gives the non-zero value.
Also we find that the large $x$-behaviors of polarized quark distributons
dramatically vary from one factorization scheme to another. Indeed, for
$x\rightarrow 1$,  the quark distributions positively diverge or
negatively diverge or remain finite,  depending on factorization schemes.
The numerical analyses performed for the parton distributions reassures the 
above
observations. From the viewpoint of
`perturbative stabilities' the OS scheme gives more appropriate
behaviors for the quark distributions than the others.
The gluon distribution  turns out to be
the same up to NLO among the six factorization schemes examined.
Furthermore, its first moment is found to be  factorization-scheme
independent up to NLO.

The same analysis on the factorization scheme dependence
of the unpolarized parton distributions of the virtual photon
can be carried out and will be discussed elsewhere.

\vspace{4cm}

\vspace{0.5cm}
\leftline{\large\bf Acknowledgement}
\vspace{0.5cm}

We thank J. Bl\"{u}mlein, S.~J.~Brodsky, J.~Kodaira,   M.~Stratmann,
O.~V.~Teryaev
and W.L.~van Neerven for valuable discussions. One of the authors
(K.S.)  would like to thank J. Bl\"{u}mlein and T.~Riemann for the hospitality
extended to him at the 5th Zeuthen Workshop ``Loops and Legs 2000".
Some of the results in this paper have been reported at the workshop.
We also thank E.~Reya and C.~Sieg for the communication regarding the
numerical evaluation of the structure function.
This work is partially supported by the
Monbusho Grant-in-Aid for Scientific Research NO.(C)(2)-12640266.

\newpage
\appendix

\noindent
{\LARGE\bf Appendix}

\section{NLO expressions for polarized parton distributions in
the virtual photon}

We give the explicit expressions of $\Delta q^\gamma_S$,
$\Delta G^\gamma$, and $\Delta q^\gamma_{NS}$ up to NLO.
They are written in
terms of  one-(two-) loop anomalous dimensions $\Delta\gamma_{ij}^{0,n}$
($\Delta\gamma_{ij}^{(1),n}$) ($i,j=\psi, G$), $\Delta\gamma_{NS}^{0,n}$
($\gamma_{NS}^{(1),n}$),  $\Delta K_l^{0,n}$ ($\Delta K_l^{(1),n}$) ($l=\psi,
G, NS$), and the one-loop photon matrix elements  of  hadronic operators,
$\Delta A_l^n$. The expressions of one-loop and $\overline {\rm MS}$
scheme-two-loop
anomalous dimensions are found, for example, in Appendix of Ref.\cite{SU}.

\subsection{Singlet quark distribution}

\bea
&&\Delta q_S^\gamma(n,Q^2,P^2)/\frac{\alpha}{8\pi\beta_0}
\nonumber\\
&&=\frac{4\pi}{\alpha_s(Q^2)}{\hat L}_S^{+n}
\left\{ 1-\left[\frac{\alpha_s(Q^2)}{\alpha_s(P^2)}\right]^
{\lambda_{+}^n/2\beta_0+1}
\right\}
+\frac{4\pi}{\alpha_s(Q^2)}{\hat L}_S^{-n}
\left\{ 1-\left[\frac{\alpha_s(Q^2)}{\alpha_s(P^2)}\right]^
{\lambda_{-}^n/2\beta_0+1}
\right\} \nonumber\\
&&+{\hat A}_S^{+n}
\left\{ 1-\left[\frac{\alpha_s(Q^2)}{\alpha_s(P^2)}\right]^
{\lambda_{+}^n/2\beta_0}
\right\}
+{\hat A}_S^{-n}
\left\{ 1-\left[\frac{\alpha_s(Q^2)}{\alpha_s(P^2)}\right]^
{\lambda_{-}^n/2\beta_0}
\right\} \nonumber\\
&&+{\hat B}_S^{+n}
\left\{ 1-\left[\frac{\alpha_s(Q^2)}{\alpha_s(P^2)}\right]^
{\lambda_{+}^n/2\beta_0+1}
\right\}
+{\hat B}_S^{-n}
\left\{ 1-\left[\frac{\alpha_s(Q^2)}{\alpha_s(P^2)}\right]^
{\lambda_{-}^n/2\beta_0+1}
\right\} \nonumber\\
&&+{\hat C}_S^{n} \label{MainSinglet}
\eea

where
\bea
{\hat L}_S^{+n}&=&\Delta K_\psi^{0,n}\cdot
\frac{\Delta
\gamma_{\psi\psi}^{0,n}-\lambda_{-}^n}{\lambda_{+}^n-\lambda_{-}^n}\cdot
\frac{1}{1+\lambda_{+}^n/2\beta_0}  \\
{\hat L}_S^{-n}&=&\Delta K_\psi^{0,n}\cdot
\frac{\Delta
\gamma_{\psi\psi}^{0,n}-\lambda_{+}^n}{\lambda_{-}^n-\lambda_{+}^n}\cdot
\frac{1}{1+\lambda_{-}^n/2\beta_0} \\
{\rm with}\qquad && \nonumber  \\
  \lambda^n_{\pm}&=&\frac{1}{2}\Bigl\{\Delta \gamma^{0,n}_{\psi\psi}+
  \Delta \gamma^{0,n}_{GG} \pm \Bigl[(\Delta \gamma^{0,n}_{\psi\psi}-
\Delta \gamma^{0,n}_{GG})^2+4 \Delta \gamma^{0,n}_{\psi G}\Delta
\gamma^{0,n}_{G\psi} \Bigr]^{1/2}\Bigr\} \\
\beta_0&=&11-2N_f/3, \qquad \beta_1=102-38N_f/3~.
\eea

and
\bea
{\hat A}_S^{+n}&=&\frac{1}
{\lambda_{+}^n(\lambda_{+}^n-\lambda_{-}^n)
(2\beta_0+\lambda_{-}^n-\lambda_{+}^n)}
\nonumber\\
&&\times\left[
\Delta K_\psi^{0,n}\left\{(\Delta
\gamma_{\psi\psi}^{0,n}-2\beta_0-\lambda_{-}^n)
\Delta \gamma_{\psi\psi}^{(1),n}
+\Delta \gamma_{G\psi}^{0,n}\Delta \gamma_{\psi G}^{(1),n}\right\}
(\Delta \gamma_{\psi\psi}^{0,n}-\lambda_{-}^n)
\right.
\nonumber\\
&&\left.+\Delta K_\psi^{0,n}\left\{(\Delta
\gamma_{\psi\psi}^{0,n}-2\beta_0-\lambda_{-}^n)
\Delta \gamma_{G\psi}^{(1),n}
+\Delta \gamma_{G\psi}^{0,n}\Delta \gamma_{GG}^{(1),n}\right\}
\Delta \gamma_{\psi G}^{0,n}\right.\nonumber\\
&&\left.+2\beta_0(2\beta_0+\lambda_{-}^n-\lambda_{+}^n)
\left\{\Delta K_{\psi}^{(1),n}(\Delta \gamma_{\psi\psi}^{0,n}-\lambda_{-}^n)
+\Delta K_{G}^{(1),n}\Delta \gamma_{\psi G}^{0,n}\right\}\right.
\nonumber\\
&&\left.
-2\beta_0(2\beta_0+\lambda_{-}^n-\lambda_{+}^n)\lambda_{+}^n\Delta A^n_{\psi}
(\Delta \gamma_{\psi\psi}^{0,n}-\lambda_{-}^n)
\right.\nonumber\\
&&\left.
-\frac{\beta_1}{\beta_0}\Delta
K_{\psi}^{0,n}(2\beta_0+\lambda_{-}^n-\lambda_{+}^n)
(2\beta_0-\lambda_{+}^n)(\Delta \gamma_{\psi\psi}^{0,n}-\lambda_{-}^n)\right]
\eea

\bea
{\hat A}_S^{-n}&=&\frac{1}
{\lambda_{-}^n(\lambda_{-}^n-\lambda_{+}^n)
(2\beta_0+\lambda_{+}^n-\lambda_{-}^n)}
\nonumber\\
&&\times\left[
\Delta K_\psi^{0,n}\left\{(\Delta
\gamma_{\psi\psi}^{0,n}-2\beta_0-\lambda_{+}^n)
\Delta \gamma_{\psi\psi}^{(1),n}
+\Delta \gamma_{G\psi}^{0,n}\Delta \gamma_{\psi G}^{(1),n}\right\}
(\Delta \gamma_{\psi\psi}^{0,n}-\lambda_{+}^n)
\right.
\nonumber\\
&&\left.+\Delta K_\psi^{0,n}\left\{(\Delta
\gamma_{\psi\psi}^{0,n}-2\beta_0-\lambda_{+}^n)
\Delta \gamma_{G\psi}^{(1),n}
+\Delta \gamma_{G\psi}^{0,n}\Delta \gamma_{GG}^{(1),n}\right\}
\Delta \gamma_{\psi G}^{0,n}\right.\nonumber\\
&&\left.+2\beta_0(2\beta_0+\lambda_{+}^n-\lambda_{-}^n)
\left\{\Delta K_{\psi}^{(1),n}(\Delta \gamma_{\psi\psi}^{0,n}-\lambda_{+}^n)
+\Delta K_{G}^{(1),n}\Delta \gamma_{\psi G}^{0,n}\right\}\right.
\nonumber\\
&&\left.
-2\beta_0(2\beta_0+\lambda_{+}^n-\lambda_{-}^n)\lambda_{-}^n\Delta A^n_{\psi}
(\Delta \gamma_{\psi\psi}^{0,n}-\lambda_{+}^n)
\right.\nonumber\\
&&\left.
-\frac{\beta_1}{\beta_0}\Delta
K_{\psi}^{0,n}(2\beta_0+\lambda_{+}^n-\lambda_{-}^n)
(2\beta_0-\lambda_{-}^n)(\Delta \gamma_{\psi\psi}^{0,n}-\lambda_{+}^n)\right]
\eea

\bea
{\hat B}_S^{+n}&=&\Delta K_{\psi}^{0,n}\cdot\frac{1}
{(2\beta_0+\lambda_{+}^n)(\lambda_{+}^n-\lambda_{-}^n)
(2\beta_0+\lambda_{+}^n-\lambda_{-}^n)}
\nonumber\\
&&\times\left[
\left\{(\Delta \gamma_{\psi\psi}^{0,n}-\lambda_{-}^n)
\Delta \gamma_{\psi\psi}^{(1),n}
+\Delta \gamma_{G\psi}^{0,n}\Delta \gamma_{\psi G}^{(1),n}\right\}
(2\beta_0+\Delta \gamma_{\psi\psi}^{0,n}-\lambda_{-}^n)
\right.
\nonumber\\
&&\left.+\left\{(\Delta \gamma_{\psi\psi}^{0,n}-\lambda_{-}^n)
\Delta \gamma_{G\psi}^{(1),n}
+\Delta \gamma_{G\psi}^{0,n}\Delta \gamma_{GG}^{(1),n}\right\}
\Delta \gamma_{\psi G}^{0,n}\right.\nonumber\\
&&\left.
-\frac{\beta_1}{\beta_0}(2\beta_0+\lambda_{+}^n-\lambda_{-}^n)
\lambda_{+}^n(\Delta \gamma_{\psi\psi}^{0,n}-\lambda_{-}^n)\right]
\eea

\bea
{\hat B}_S^{-n}&=& \Delta K_{\psi}^{0,n}\cdot\frac{1}
{(2\beta_0+\lambda_{-}^n)(\lambda_{-}^n-\lambda_{+}^n)
(2\beta_0+\lambda_{-}^n-\lambda_{+}^n)}
\nonumber\\
&&\times\left[
\left\{(\Delta \gamma_{\psi\psi}^{0,n}-\lambda_{+}^n)
\Delta \gamma_{\psi\psi}^{(1),n}
+\Delta \gamma_{G\psi}^{0,n}\Delta \gamma_{\psi G}^{(1),n}\right\}
(2\beta_0+\Delta \gamma_{\psi\psi}^{0,n}-\lambda_{+}^n)
\right.
\nonumber\\
&&\left.+\left\{(\Delta \gamma_{\psi\psi}^{0,n}-\lambda_{+}^n)
\Delta \gamma_{G\psi}^{(1),n}
+\Delta \gamma_{G\psi}^{0,n}\Delta \gamma_{GG}^{(1),n}\right\}
\Delta \gamma_{\psi G}^{0,n}\right.\nonumber\\
&&\left.
-\frac{\beta_1}{\beta_0}(2\beta_0+\lambda_{-}^n-\lambda_{+}^n)
\lambda_{-}^n(\Delta \gamma_{\psi\psi}^{0,n}-\lambda_{+}^n)\right]
\nonumber\\
&&\nonumber\\
&&\nonumber\\
{\hat C}_S^{n}&=&2\beta_0\Delta A^n_{\psi}
\eea

\bigskip
\subsection{Gluon distribution}

\bea
&&\Delta G^\gamma(n,Q^2,P^2)/\frac{\alpha}{8\pi\beta_0}
\nonumber\\
&&=\frac{4\pi}{\alpha_s(Q^2)}{\hat L}_G^{+n}
\left\{ 1-\left[\frac{\alpha_s(Q^2)}{\alpha_s(P^2)}\right]^
{\lambda_{+}^n/2\beta_0+1}
\right\}
+\frac{4\pi}{\alpha_s(Q^2)}{\hat L}_G^{-n}
\left\{ 1-\left[\frac{\alpha_s(Q^2)}{\alpha_s(P^2)}\right]^
{\lambda_{-}^n/2\beta_0+1}
\right\} \nonumber\\
&&+{\hat A}_G^{+n}
\left\{ 1-\left[\frac{\alpha_s(Q^2)}{\alpha_s(P^2)}\right]^
{\lambda_{+}^n/2\beta_0}
\right\}
+{\hat A}_G^{-n}
\left\{ 1-\left[\frac{\alpha_s(Q^2)}{\alpha_s(P^2)}\right]^
{\lambda_{-}^n/2\beta_0}
\right\} \nonumber\\
&&+{\hat B}_G^{+n}
\left\{ 1-\left[\frac{\alpha_s(Q^2)}{\alpha_s(P^2)}\right]^
{\lambda_{+}^n/2\beta_0+1}
\right\}
+{\hat B}_G^{-n}
\left\{ 1-\left[\frac{\alpha_s(Q^2)}{\alpha_s(P^2)}\right]^
{\lambda_{-}^n/2\beta_0+1}
\right\}
\eea
where
\bea
&&{\hat L}_G^{+n}=\frac{\Delta K_\psi^{0,n}\Delta \gamma_{G\psi}^{0,n}}
{\lambda_{+}^n-\lambda_{-}^n}
\cdot\frac{1}{1+\lambda_{+}^n/2\beta_0}  \\
&&{\hat L}_G^{-n}=\frac{\Delta K_\psi^{0,n}\Delta \gamma_{G\psi}^{0,n}}
{\lambda_{-}^n-\lambda_{+}^n}
\cdot\frac{1}{1+\lambda_{-}^n/2\beta_0}
\eea
and
\bea
{\hat A}_G^{+n}&=&\frac{1}
{\lambda_{+}^n(\lambda_{+}^n-\lambda_{-}^n)
(2\beta_0+\lambda_{-}^n-\lambda_{+}^n)}
\nonumber\\
&&\times\left[
\Delta K_\psi^{0,n}\left\{(\Delta
\gamma_{\psi\psi}^{0,n}-2\beta_0-\lambda_{-}^n)
\Delta \gamma_{\psi\psi}^{(1),n}
+\Delta \gamma_{G\psi}^{0,n}\Delta \gamma_{\psi G}^{(1),n}\right\}\Delta
\gamma_{G\psi}^{0,n}
\right.
\nonumber\\
&&\left.+\Delta K_\psi^{0,n}\left\{(\Delta
\gamma_{\psi\psi}^{0,n}-2\beta_0-\lambda_{-}^n)
\Delta \gamma_{G\psi}^{(1),n}
+\Delta \gamma_{G\psi}^{0,n}\Delta \gamma_{GG}^{(1),n}\right\}
(\Delta \gamma_{GG}^{0,n}-\lambda_{-}^n)\right.\nonumber\\
&&\left.+2\beta_0(2\beta_0+\lambda_{-}^n-\lambda_{+}^n)
\left\{\Delta K_{\psi}^{(1),n}\Delta \gamma_{G\psi}^{0,n}+\Delta
K_{G}^{(1),n}(\Delta \gamma_{GG}^{0,n} -\lambda_{-}^n)\right\}\right.
\nonumber\\
&&\left.
-2\beta_0(2\beta_0+\lambda_{-}^n-\lambda_{+}^n)\lambda_{+}^n\Delta A^n_{\psi}
\Delta \gamma_{G\psi}^{0,n}
\right.\nonumber\\
&&\left.
-\frac{\beta_1}{\beta_0}\Delta
K_{\psi}^{0,n}(2\beta_0+\lambda_{-}^n-\lambda_{+}^n)
(2\beta_0-\lambda_{+}^n)\Delta \gamma_{G\psi}^{0,n}\right]
\eea

\bea
{\hat A}_G^{-n}&=&\frac{1}
{\lambda_{-}^n(\lambda_{-}^n-\lambda_{+}^n)
(2\beta_0+\lambda_{+}^n-\lambda_{-}^n)}
\nonumber\\
&&\times\left[
\Delta K_\psi^{0,n}\left\{(\Delta
\gamma_{\psi\psi}^{0,n}-2\beta_0-\lambda_{+}^n)
\Delta \gamma_{\psi\psi}^{(1),n}
+\Delta \gamma_{G\psi}^{0,n}\Delta \gamma_{\psi G}^{(1),n}\right\}\Delta
\gamma_{G\psi}^{0,n}
\right.
\nonumber\\
&&\left.+\Delta K_\psi^{0,n}\left\{(\Delta
\gamma_{\psi\psi}^{0,n}-2\beta_0-\lambda_{+}^n)
\Delta \gamma_{G\psi}^{(1),n}
+\Delta \gamma_{G\psi}^{0,n}\Delta \gamma_{GG}^{(1),n}\right\}
(\Delta \gamma_{GG}^{0,n}-\lambda_{+}^n)\right.\nonumber\\
&&\left.+2\beta_0(2\beta_0+\lambda_{+}^n-\lambda_{-}^n)
\left\{\Delta K_{\psi}^{(1),n}\Delta \gamma_{G\psi}^{0,n}+\Delta
K_{G}^{(1),n}(\Delta \gamma_{GG}^{0,n} -\lambda_{+}^n)\right\}\right.
\nonumber\\
&&\left.
-2\beta_0(2\beta_0+\lambda_{+}^n-\lambda_{-}^n)\lambda_{-}^n\Delta A^n_{\psi}
\Delta \gamma_{G\psi}^{0,n}
\right.\nonumber\\
&&\left.
-\frac{\beta_1}{\beta_0}\Delta
K_{\psi}^{0,n}(2\beta_0+\lambda_{+}^n-\lambda_{-}^n)
(2\beta_0-\lambda_{-}^n)\Delta \gamma_{G\psi}^{0,n}\right]
\eea

\bea
{\hat B}_G^{+n}&=& \Delta K_{\psi}^{0,n}\cdot\frac{1}
{(2\beta_0+\lambda_{+}^n)(\lambda_{+}^n-\lambda_{-}^n)
(2\beta_0+\lambda_{+}^n-\lambda_{-}^n)}
\nonumber\\
&&\times\left[
\left\{(\Delta \gamma_{\psi\psi}^{0,n}-\lambda_{-}^n)
\Delta \gamma_{G\psi}^{(1),n}
+\Delta \gamma_{G\psi}^{0,n}\Delta \gamma_{GG}^{(1),n}\right\}
(2\beta_0+\Delta \gamma_{GG}^{0,n}-\lambda_{-}^n)
\right.
\nonumber\\
&&\left.+\left\{(\Delta \gamma_{\psi\psi}^{0,n}-\lambda_{-}^n)
\Delta \gamma_{\psi\psi}^{(1),n}
+\Delta \gamma_{G\psi}^{0,n}\Delta \gamma_{\psi G}^{(1),n}\right\}
\Delta \gamma_{G\psi}^{0,n}\right.\nonumber\\
&&\left.
-\frac{\beta_1}{\beta_0}(2\beta_0+\lambda_{+}^n-\lambda_{-}^n)
\lambda_{+}^n\Delta \gamma_{G\psi}^{0,n}\right]
\eea

\bea
{\hat B}_G^{-n}&=& \Delta K_{\psi}^{0,n}\cdot\frac{1}
{(2\beta_0+\lambda_{-}^n)(\lambda_{-}^n-\lambda_{+}^n)
(2\beta_0+\lambda_{-}^n-\lambda_{+}^n)}
\nonumber\\
&&\times\left[
\left\{(\Delta \gamma_{\psi\psi}^{0,n}-\lambda_{+}^n)
\Delta \gamma_{G\psi}^{(1),n}
+\Delta \gamma_{G\psi}^{0,n}\Delta \gamma_{GG}^{(1),n}\right\}
(2\beta_0+\Delta \gamma_{GG}^{0,n}-\lambda_{+}^n)
\right.
\nonumber\\
&&\left.+\left\{(\Delta \gamma_{\psi\psi}^{0,n}-\lambda_{+}^n)
\Delta \gamma_{\psi\psi}^{(1),n}
+\Delta \gamma_{G\psi}^{0,n}\Delta \gamma_{\psi G}^{(1),n}\right\}
\Delta \gamma_{G\psi}^{0,n}\right.\nonumber\\
&&\left.
-\frac{\beta_1}{\beta_0}(2\beta_0+\lambda_{-}^n-\lambda_{+}^n)
\lambda_{-}^n\Delta \gamma_{G\psi}^{0,n}\right]
\eea

\subsection{Non-singlet quark}

\bea
\Delta q_{NS}^\gamma(n,Q^2,P^2)/\frac{\alpha}{8\pi\beta_0}
&=&\frac{4\pi}{\alpha_s(Q^2)}{\hat L}_{NS}^{n}
\left\{ 1-\left[\frac{\alpha_s(Q^2)}{\alpha_s(P^2)}\right]^
{\lambda_{NS}^n/2\beta_0+1} \right\}  \nonumber\\
&&+{\hat A}_{NS}^{n}
\left\{ 1-\left[\frac{\alpha_s(Q^2)}{\alpha_s(P^2)}\right]^
{\lambda_{NS}^n/2\beta_0} \right\}  \nonumber\\
&&+{\hat B}_{NS}^{n}
\left\{ 1-\left[\frac{\alpha_s(Q^2)}{\alpha_s(P^2)}\right]^
{\lambda_{NS}^n/2\beta_0+1} \right\}  \nonumber\\
&&+{\hat C}_{NS}^{n}   \label{MainNonsinglet}
\eea
where
\bea
{\hat L}_{NS}^{n}&=&\Delta K_{NS}^{0,n}\cdot
\frac{1}{1+\lambda_{NS}^n/2\beta_0} \\
{\hat A}_{NS}^{n}&=&
\frac{1}{\lambda_{NS}^n}\biggl\{-\Delta K_{NS}^{0,n}\Delta
\gamma_{NS}^{(1),n}+2\beta_0
\Delta K_{NS}^{(1),n}-2\beta_0\lambda_{NS}^n \Delta A^n_{NS} \nonumber\\
& &\qquad \qquad -\frac{\beta_1}{\beta_0}\Delta
K_{NS}^{0,n}(2\beta_0-\lambda_{NS}^n) \biggr\} \\
{\hat B}_{NS}^{n}&=&
\Delta K_{NS}^{0,n}\frac{1}{2\beta_0+\lambda_{NS}^n}\left(
\Delta \gamma_{NS}^{(1),n}-\frac{\beta_1}{\beta_0}\lambda_{NS}^n\right) \\
{\hat C}_{NS}^{n}&=&2\beta_0 \Delta A^n_{NS} \\
{\rm with}\qquad &&\nonumber \\
  \lambda^n_{NS}&=&\Delta\gamma^{0,n}_{NS}
\eea

\bigskip

\section{The first moments}

\subsection{One-loop order}
\bea
     \Delta\gamma^{0,n=1}_{NS}&=&\Delta\gamma^{0,n=1}_{\psi\psi}=0
\label{gamma0qq}\\
     \Delta\gamma^{0,n=1}_{\psi G}&=&0,  \qquad  \qquad
     \Delta\gamma^{0,n=1}_{G \psi}=-6C_F \\
     \Delta\gamma^{0,n=1}_{G G}&=& -\frac{22}{3}C_A +\frac{8}{3}T_f
=-2\beta_0   \\
     \lambda_+^{n=1}&=&0,  \qquad  \qquad \lambda_-^{n=1}=-2\beta_0  \\
    \Delta K^{0,n=1}_{NS} &=& \Delta K^{0,n=1}_{\psi}=0  \label{K0SNS}
\eea
where
\be
C_A=3~,\qquad C_F=\frac{4}{3}~,\qquad T_f=\frac{N_f}{2}
\ee
with $N_f$ being the number of flavors.

\subsection{$\overline {\rm MS}$ scheme}
\bea
   \Delta\gamma^{(1),n=1}_{NS,~\overline {\rm MS}}&=& 0  \\
  & & \nonumber  \\
   \Delta\gamma^{(1),n=1}_{\psi\psi,~\overline {\rm MS}}&=& 24 C_F T_f   \\
   \Delta\gamma^{(1),n=1}_{\psi G,~\overline {\rm MS}}&=& 0  \\
  \Delta \gamma^{(1),n=1}_{G \psi,~\overline {\rm MS}}&=&
18C_F^2-\frac{142}{3}C_AC_F+\frac{8}{3}C_FT_f  \\
\Delta\gamma^{(1),n=1}_{GG,~\overline {\rm MS}}&=&8C_F T_f +\frac{40}{3}C_A T_f
-\frac{68}{3}C_A^2  =-2\beta_1 \\
   & & \nonumber  \\
   \Delta K^{(1),n=1}_{\psi,~\overline {\rm MS}} &=&\Delta
K^{(1),n=1}_{G,~\overline
{\rm MS}}=
\Delta K^{(1),n=1}_{NS,~\overline {\rm MS}}=0  \\
   & & \nonumber  \\
\Delta A^{n=1}_{\psi,~\overline {\rm MS}}&=&-12\langle e^2 \rangle N_f  \\
\Delta A^{n=1}_{G,~\overline {\rm MS}}&=&0   \\
\Delta A^{n=1}_{NS,~\overline {\rm MS}}&=&-12(\langle e^4\rangle-\langle e^2
\rangle^2)
N_f
\\
   & & \nonumber  \\
\Delta B^{n=1}_{\psi,~\overline {\rm MS}}&=&\Delta B^{n=1}_{NS,~\overline
{\rm MS}}=-3C_F
\\
\Delta B^{n=1}_{G,~\overline {\rm MS}}&=&\frac{N_f}{2}\Delta
B^{n=1}_{\gamma,~\overline
{\rm MS}}=0 \\
\eea

\subsection{CI-like schemes (CI, AB, OS, AR)}
\bea
  \Delta \gamma^{(1),n=1}_{NS,~a}&=& 0  \\
  & & \nonumber  \\
   \Delta\gamma^{(1),n=1}_{\psi\psi,~a}&=& 0  \\
   \Delta\gamma^{(1),n=1}_{\psi G,~a}&=& \Delta\gamma^{(1),n=1}_{\psi
G,~\overline
{\rm MS}}=0
\\
  \Delta \gamma^{(1),n=1}_{G \psi,~a}&=&\Delta \gamma^{(1),n=1}_{G
\psi,~\overline
{\rm MS}}= 18C_F^2-\frac{142}{3}C_AC_F+\frac{8}{3}C_FT_f  \\
\Delta\gamma^{(1),n=1}_{GG,~a}&=&32C_F T_f +\frac{40}{3}C_A T_f
-\frac{68}{3}C_A^2  =-2\beta_1 +12N_f C_F\\
   & & \nonumber  \\
  \Delta  K^{(1),n=1}_{\psi,~a} &=&\Delta K^{(1),n=1}_{NS,~a}=0  \\
\Delta K^{(1),n=1}_{G,~a}&=&-72\langle e^2 \rangle N_f C_F
\label{DeltaKGCI}\\
   & & \nonumber  \\
\Delta A^{n=1}_{\psi,~a}&=&\Delta A^{n=1}_{G,~a}=\Delta A^{n=1}_{NS,~a}=0 \\
   & & \nonumber  \\
\Delta B^{n=1}_{\psi,~a}&=&\Delta B^{n=1}_{NS,~a}=-3C_F \\
\Delta B^{n=1}_{G,~a}&=&\frac{N_f}{2}\Delta B^{n=1}_{\gamma,~a}=-2N_f
\eea

\bigskip

\section{Derivation of Eqs.(\ref{QSMS}) and (\ref{QNSMS}) }

We observe that, in the $\overline {\rm MS}$ scheme, we have
$\Delta \Kv^{0,n=1}=\Delta \Kv^{(1),n=1}=0$,
where $\Delta \Kv^n=(\Delta K^n_{\psi},\Delta K^n_{G} ,\Delta K^n_{NS})$.\ \
(Note $\Delta K^{(1),n=1}_{G,~\rm CI-like}\neq 0$. See Eq.(\ref{DeltaKGCI}). )
Then, up to  NLO, the parton distributions
$\Delta \qv^{\gamma}(n=1)\vert_{\overline {\rm MS}}=(\Delta q_S^{\gamma},~
\Delta G^{\gamma},~ \Delta q_{NS}^{\gamma})\vert_{\overline {\rm MS}}$~
satisfy
a  homogeneous differential equation instead of an  inhomogenious one:
\be
\frac{d~\Delta\qv^\gamma(n=1,Q^2,P^2)\vert_{\overline {\rm MS}}}{d\ln Q^2}=
\Delta\qv^\gamma(n=1,Q^2,P^2)\vert_{\overline {\rm MS}}~\Delta
P(n=1,Q^2)\vert_{\overline {\rm MS}}  \label{Homogeneous}
\label{Appendixevol}
\ee
where the $3\times 3$ splitting function matrix $\Delta P$ is the hadronic
part of
$\Delta \widetilde{P}$ given in Eq.(\ref{4by4matrix}).
Expanding $\Delta P(n=1,Q^2)\vert_{\overline {\rm MS}}$ as
\be
\Delta  P(n=1,Q^2)\vert_{\overline {\rm MS}}=\frac{\alpha_s(Q^2)}{2\pi}\Delta
P^{(0)}_{n=1}+
\Bigl[\frac{\alpha_s(Q^2)}{2\pi}\Bigr]^2 \Delta P^{(1)}_{n=1}
\vert_{\overline {\rm MS}}+ \cdots ~,
\ee
and introducing $t$ instead of $Q^2$ as the evolution variable
\be
  t \equiv \frac{2}{\beta_0}{\rm ln}\frac{\alpha_s(P^2)}{\alpha_s(Q^2)},
\ee
we find that Eq.(\ref{Homogeneous}) is rewritten as
\be
\frac{d~\Delta\qv^\gamma_{n=1}(t)\vert_{\overline {\rm MS}}}{dt}=
\Delta\qv^\gamma_{n=1}(t)\vert_{\overline {\rm MS}}\left\{\Delta
P^{(0)}_{n=1}+\frac{\alpha_s(t)}{2\pi}\Bigl[ \Delta
P^{(1)}_{n=1}\vert_{\overline
{\rm MS}}-\frac{\beta_1}{2\beta_0}\Delta P^{(0)}_{n=1}   \Bigr]+ {\cal
O}(\alpha_s^2)
\right\}~.
\ee

We look for the solution in the following form:
\be
\Delta\qv^\gamma_{n=1}(t)\vert_{\overline {\rm MS}}=
\Delta\qv^{\gamma(0)}_{n=1}(t)+\Delta\qv^{\gamma(1)}_{n=1}(t)
\vert_{\overline {\rm MS}}
\ee
with the initial condition (see Eq.(\ref{PhotonMSn})),
\bea
\Delta\qv^{\gamma(0)}_{n=1}(0)&=&0   \label{InitialLO}\\
\Delta\qv^{\gamma(1)}_{n=1}(0)\vert_{\overline
{\rm MS}}&=&\frac{\alpha}{4\pi}{\Delta \Av_{n=1}}\vert_{\overline
{\rm MS}} \nonumber  \\ &=&-\frac{3\alpha}{\pi}N_f
\Bigl(\langle e^2 \rangle~,~0~,~\langle e^4\rangle-\langle e^2 \rangle^2
\Bigr) \label{CNL}
\eea
In the LO, we easily find that $\Delta\qv^{\gamma(0)}_{n=1}(t)=0$ due to the
initial condition (\ref{InitialLO}).

The evolution equation in the NLO is written as
\be
\frac{d~\Delta\qv^{\gamma(1)}_{n=1}(t)\vert_{\overline {\rm MS}}}{dt}=
\Delta\qv^{\gamma(1)}_{n=1}(t)\vert_{\overline {\rm MS}}
\left\{\Delta
P^{(0)}_{n=1}+\frac{\alpha_s(t)}{2\pi}\Bigl[ \Delta
P^{(1)}_{n=1}\vert_{\overline
{\rm MS}}-\frac{\beta_1}{2\beta_0}\Delta P^{(0)}_{n=1}   \Bigr]
\right\}~,
\ee
and we obtain for the solution
\be
  \Delta\qv^{\gamma(1)}_{n=1}(t)\vert_{\overline {\rm MS}}=
\Delta\qv^{\gamma(1)}_{n=1}(0)\vert_{\overline {\rm MS}} ~
{\rm exp}~\Bigl( M \Bigr),~ \label{NLOQuarkSolution}
\ee
where
\be
M=\Delta P^{(0)}_{n=1}~t+\frac{1}{\beta_0}
\Bigl[ \frac{\alpha_s(0)}{\pi}-\frac{\alpha_s(t)}{\pi}  \Bigr]
\Bigl[ \Delta P^{(1)}_{n=1}\vert_{\overline
{\rm MS}}-\frac{\beta_1}{2\beta_0}\Delta P^{(0)}_{n=1}\Bigr]
\ee
Since
\be
\Delta P^{(0)}_{n=1}=-\frac{1}{4}\Delta{\hat \gamma}^{0}_{n=1}~,
\qquad \Delta P^{(1)}_{n=1}\vert_{\overline {\rm MS}}=
-\frac{1}{8}\Delta{\hat \gamma}^{(1)}_{n=1}\vert_{\overline {\rm MS}}~,
\ee
and using the information on the first moments of anomalous dimensions
which are listed in Appendices B.1 and B.2, we find that $M$ turns out
to be a triangular matrix in the following form:
\be
M= \left(\matrix{a&b&0\cr
0&c&0 \cr 0&0&d } \right)
\ee
with
\bea
a&=&\frac{1}{\beta_0}
\Bigl[ \frac{\alpha_s(0)}{\pi}-\frac{\alpha_s(t)}{\pi}  \Bigr]
\left(-\frac{1}{8} \gamma_{\psi\psi, ~\overline {\rm MS}}^{(1),n=1}  \right)
\label{aSingQuark}\\
b&=&\frac{3}{2}C_Ft+\frac{1}{\beta_0}
\Bigl[ \frac{\alpha_s(0)}{\pi}-\frac{\alpha_s(t)}{\pi}  \Bigr]
\left(-\frac{1}{8} \gamma_{G\psi, ~\overline {\rm MS}}^{(1),n=1} -
\frac{3\beta_1}{4\beta_0}C_F \right) \\
c&=&\frac{1}{2}\beta_0 t  \\
d&=&\frac{1}{\beta_0}
\Bigl[ \frac{\alpha_s(0)}{\pi}-\frac{\alpha_s(t)}{\pi}  \Bigr]
\left(-\frac{1}{8} \gamma_{NS, ~\overline {\rm MS}}^{(1),n=1}  \right)
\label{dNonSingQuark}
\eea
The matrix ${\rm exp}(M)$ is, therefore, written in the form
\be
{\rm exp}(M)=\left(\matrix{e^a&B&0\cr
0&e^c&0 \cr 0&0&e^d } \right)  \label{triangularmatrix}
\ee
and thus we obtain from Eqs.(\ref{CNL}) and (\ref{NLOQuarkSolution}),
\bea
\Delta q_S^{\gamma}(n=1,Q^2,P^2)\vert_{\overline {\rm MS}}&=&
-\frac{3\alpha}{\pi}N_f <e^2>
{\rm exp}\left\{- \frac{1}{8\beta_0}
\Bigl[ \frac{\alpha_s(0)}{\pi}-\frac{\alpha_s(t)}{\pi}  \Bigr]
\Delta \gamma_{\psi\psi, ~\overline {\rm MS}}^{(1),n=1} \right\} \nonumber  \\
&\approx&-\frac{3\alpha}{\pi}N_f <e^2>
\left\{1-\frac{2}{\beta_0}\Bigl[\frac{\alpha_s(P^2)}{\pi}
-\frac{\alpha_s(Q^2)}{\pi}\Bigr] N_f\right\} \nonumber \\
& &  \\
\Delta q_{NS}^{\gamma}(n=1,Q^2,P^2)\vert_{\overline {\rm MS}}&=&
-\frac{3\alpha}{\pi}N_f(<e^4>-<e>^2 ) \nonumber  \\
& &\qquad \qquad \times
{\rm exp}\left\{- \frac{1}{8\beta_0}
\Bigl[ \frac{\alpha_s(0)}{\pi}-\frac{\alpha_s(t)}{\pi}  \Bigr]
\Delta \gamma_{NS, ~\overline {\rm MS}}^{(1),n=1}  \right\} \nonumber  \\
&=&-\frac{3\alpha}{\pi}N_f(<e^4>-<e>^2 )
\eea
where in the last line we use the fact $\Delta \gamma^{(1),n=1}_{NS,~\overline
{\rm MS}}=0$.

\bigskip

Incidentally, under the following approximtion,
\be
b\approx \frac{3}{2}C_Ft~, \qquad    a+c\approx c ~,
\ee
$B$ is evaluated as
\bea
B&\approx& b\left\{1+\frac{1}{2}c+\frac{1}{3!}c^2
  +\frac{1}{4!}c^3 +\cdots
\right\}
=\frac{b}{c}\Bigl[e^c-1   \Bigr] \nonumber \\
&\approx&\frac{3C_F}{\beta_0}\left[\frac{\alpha_s(P^2)}{\alpha_s(Q^2)} -1
  \right]~. \label{Bapprox}
\eea
This leads to the expression for the first moment of gluon distribution
$\Delta G^{\gamma}(n=1,Q^2,P^2)\vert_{\overline {\rm MS}}$ given in
Eq.(\ref{n=1Gluon}).

\newpage

\newpage
\vspace{2cm}
\noindent
{\large Figure Captions}
\baselineskip 16pt

\begin{enumerate}

\item[Fig. 1] \quad
Deep inelastic scattering on a polarized virtual photon
in polarized $e^{+}e^{-}$ collision,
$e^{+}e^{-} \rightarrow  e^{+}e^{-} + $ hadrons (quarks and gluons).
The arrows indicate the polarizations of the $e^{+}$, $e^{-}$ and
virtual photons.
The mass squared of the \lq probe\rq \ (\lq target\rq)
photon is $-Q^2(-P^2)$ ($\Lambda^2 \ll P^2 \ll Q^2$).

\item[Fig. 2] \quad
Factorization scheme dependence of the polarized
singlet quark distribution
$\Delta q_S^{\gamma}(x,Q^2,P^2)$ up to  NLO in units of
$(3N_f \langle e^2\rangle \alpha/\pi){\rm ln}(Q^2/P^2)$ with $N_f=3$,
$Q^2=30~{\rm GeV}^2$, $P^2=1~ {\rm GeV}^2$, and
the QCD scale parameter
$\Lambda=0.2~{\rm GeV}$,
for $\overline {\rm MS}$ (dash-dotted line), CI (solid line),
AB (short-dashed line),  OS (long-dashed line) ,
AR (dashed line) and ${\rm DIS}_\gamma$ (dash-2dotted line) schemes.

\item[Fig. 3]\quad
The polarized singlet quark distribution
$\Delta q_S^{\gamma}(x,Q^2,P^2)$ up to  NLO predicted
by the OS and ${\rm DIS}_\gamma$ schemes  in units of
$(3N_f \langle e^2\rangle \alpha/\pi){\rm ln}(Q^2/P^2)$ for $N_f=3$,
$Q^2=30~{\rm GeV}^2$, $P^2=1~ {\rm GeV}^2$, and
$\Lambda=0.2~{\rm GeV}$, together with the LO result.

\item[Fig. 4]\quad
The polarized singlet quark distribution
$\Delta q_S^{\gamma}(x,Q^2,P^2)$ up to  NLO
in the OS scheme in units of $(3 N_f \langle e^2\rangle \alpha
/\pi){\rm ln}(Q^2/P^2)$ with three different $Q^2$ values,
for $N_f=3$, $P^2=1~ {\rm GeV}^2$, and $\Lambda=0.2~ {\rm GeV}$.

\item[Fig. 5]\quad
The polarized gluon distribution $\Delta G^{\gamma}(x,Q^2,P^2)$ beyond the LO
in units of $(3 N_f \langle e^2\rangle \alpha /\pi){\rm ln}(Q^2/P^2)$  with
three
different $Q^2$ values, for $N_f=3$, $P^2=1~ {\rm GeV}^2$, and
$\Lambda=0.2~ {\rm GeV}$.

\item[Fig. 6]\quad
Polarized virtual photon structure function $g_1^\gamma(x,Q^2,P^2)$
up to NLO in units of $(3N_f\alpha \langle e^4\rangle /\pi)
\ln(Q^2/P^2)$ for $Q^2=30$ GeV$^2$, and $P^2=1$ GeV$^2$
and the QCD scale parameter $\Lambda=0.2$ GeV with $N_f=3$ (solid line).
We also plot the LO result (long-dashed line),
the Box (tree) diagram (2dash-dotted line) and the Box including
non-leading contribution, Box (NL) (short-dashed line).

\item[Fig. 7]\quad
Point-like piece of the real photon structure function
$g_1^\gamma(x,Q^2)$ in NLO in units of $(3N_f\alpha \langle e^4\rangle /\pi)
\ln(Q^2/\Lambda^2)$
for $Q^2=30$ GeV$^2$ with  $\Lambda=0.2$ GeV, $N_f=3$ (solid line).
Also plotted are the LO result (long-dashed line) and the Box (tree)
diagram contribution (short-dashed line).

\end{enumerate}


\newpage
\pagestyle{empty}
\input epsf.sty
\begin{figure}
\centerline{
\epsfxsize=11cm
\epsfbox{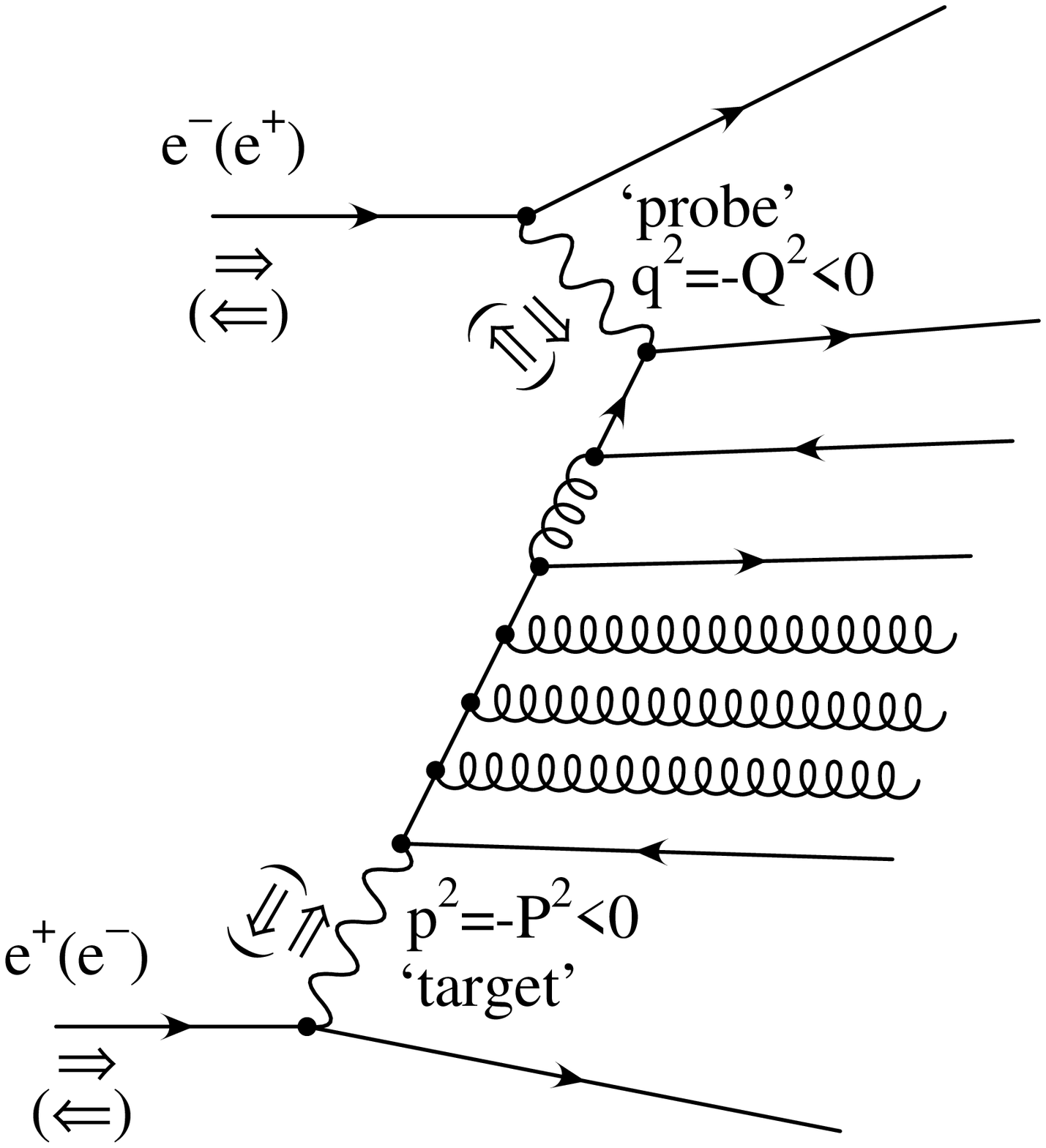}}
\vspace{-1.5cm}
\vspace{1cm}
\centerline{\large\bf Fig. 1}
\end{figure}

\newpage
\pagestyle{empty}
\input epsf.sty
\begin{figure}
\centerline{
\epsfxsize=16cm
\epsfbox{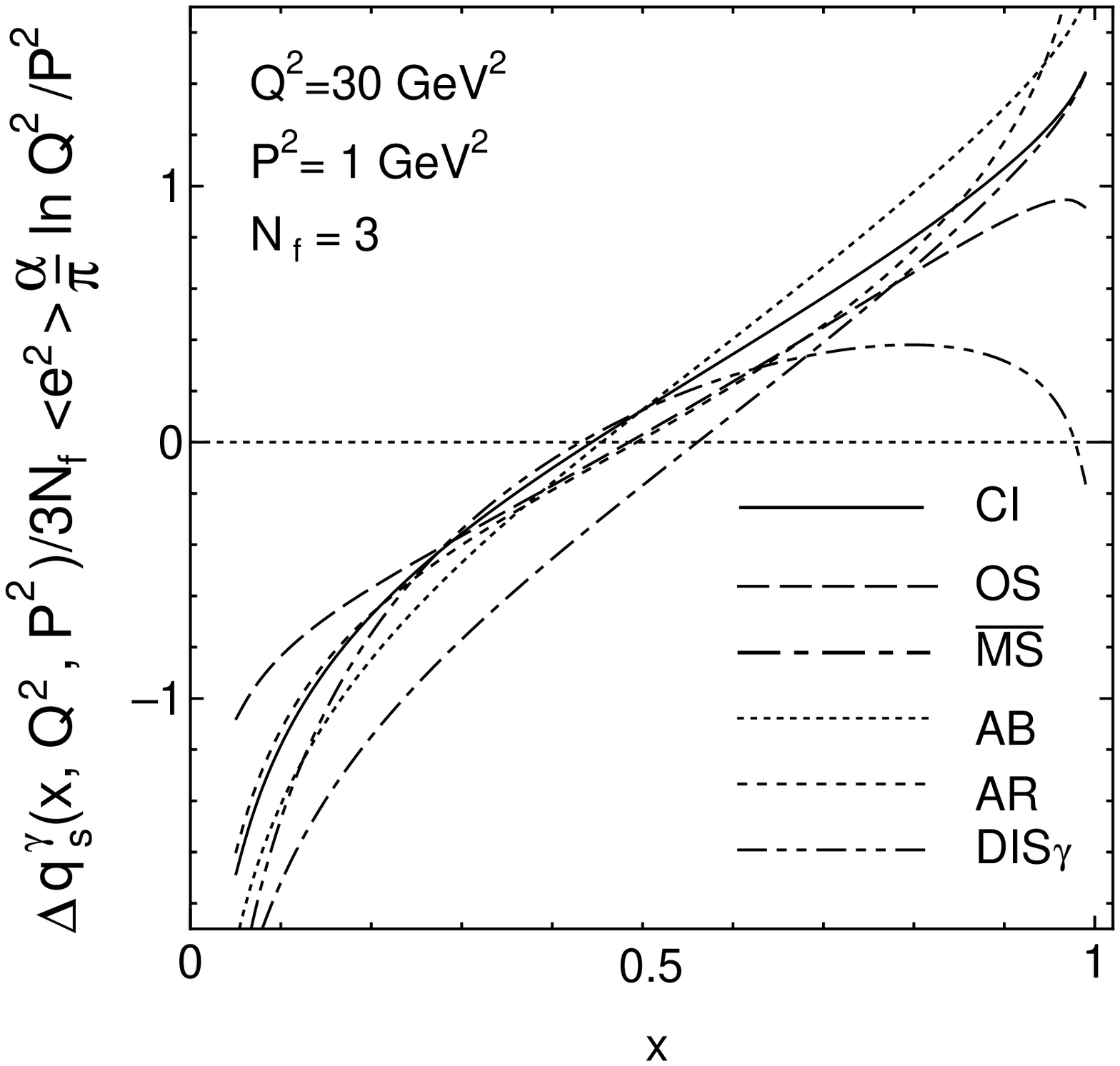}}
\vspace{-3.5cm}
\centerline{\large\bf Fig. 2}
\end{figure}

\newpage
\pagestyle{empty}
\input epsf.sty
\begin{figure}
\centerline{
\epsfxsize=16cm
\epsfbox{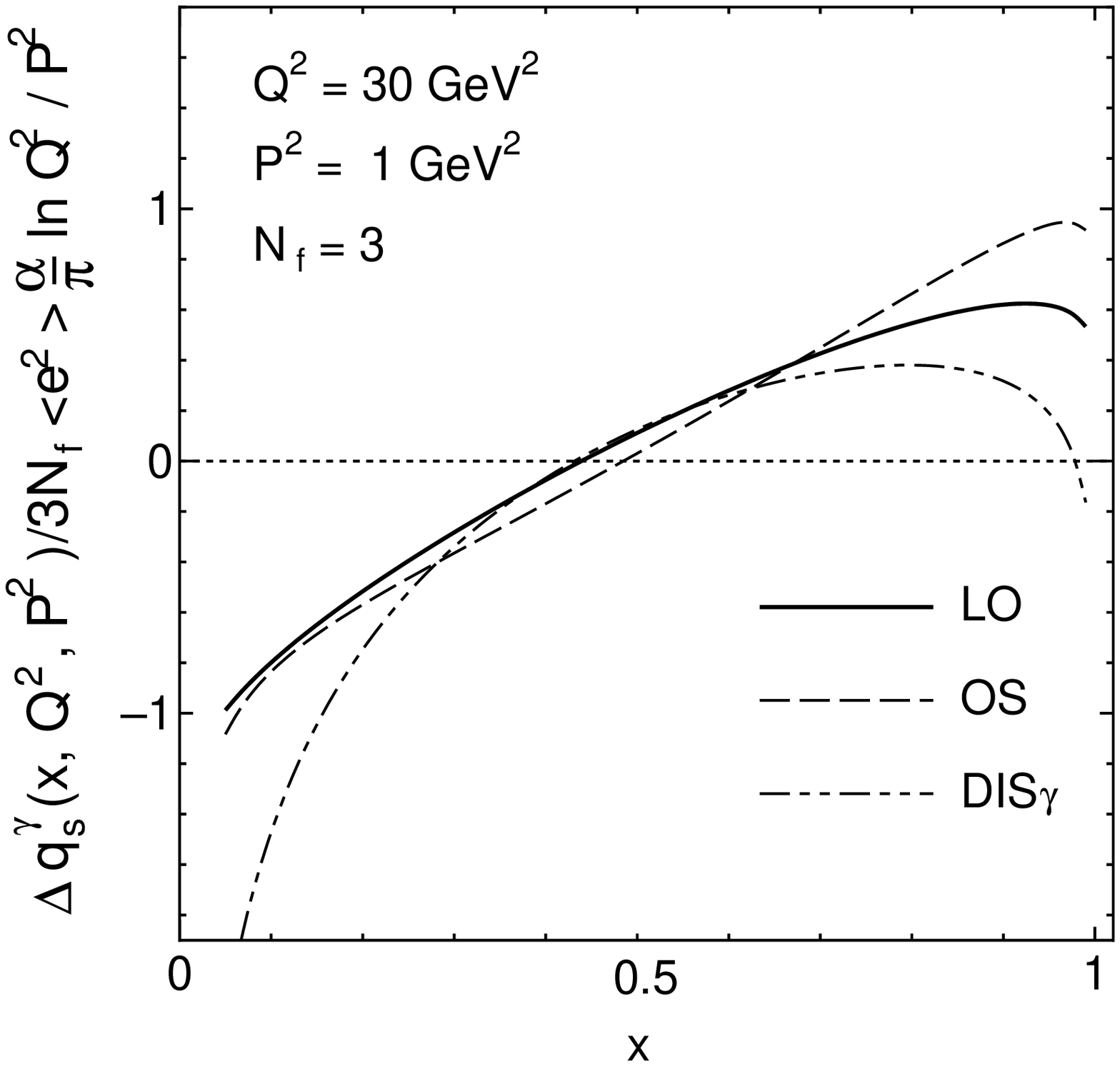}}
\vspace{-3.5cm}
\centerline{\large\bf Fig. 3}
\end{figure}

\newpage
\pagestyle{empty}
\input epsf.sty
\begin{figure}
\centerline{
\epsfxsize=16cm
\epsfbox{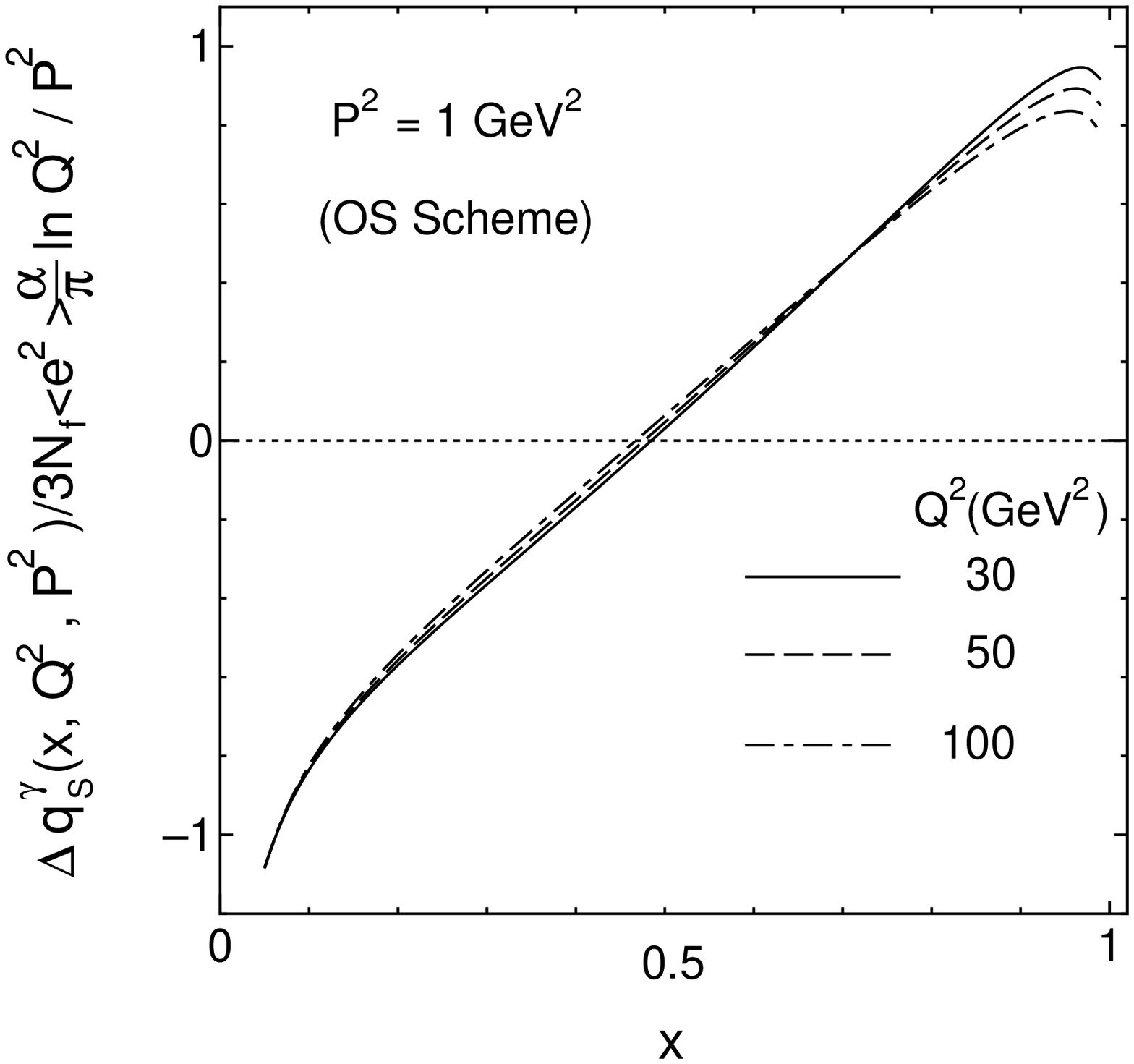}}
\vspace{-3.5cm}
\centerline{\large\bf Fig. 4}
\end{figure}

\newpage
\pagestyle{empty}
\input epsf.sty
\begin{figure}
\centerline{
\epsfxsize=16cm
\epsfbox{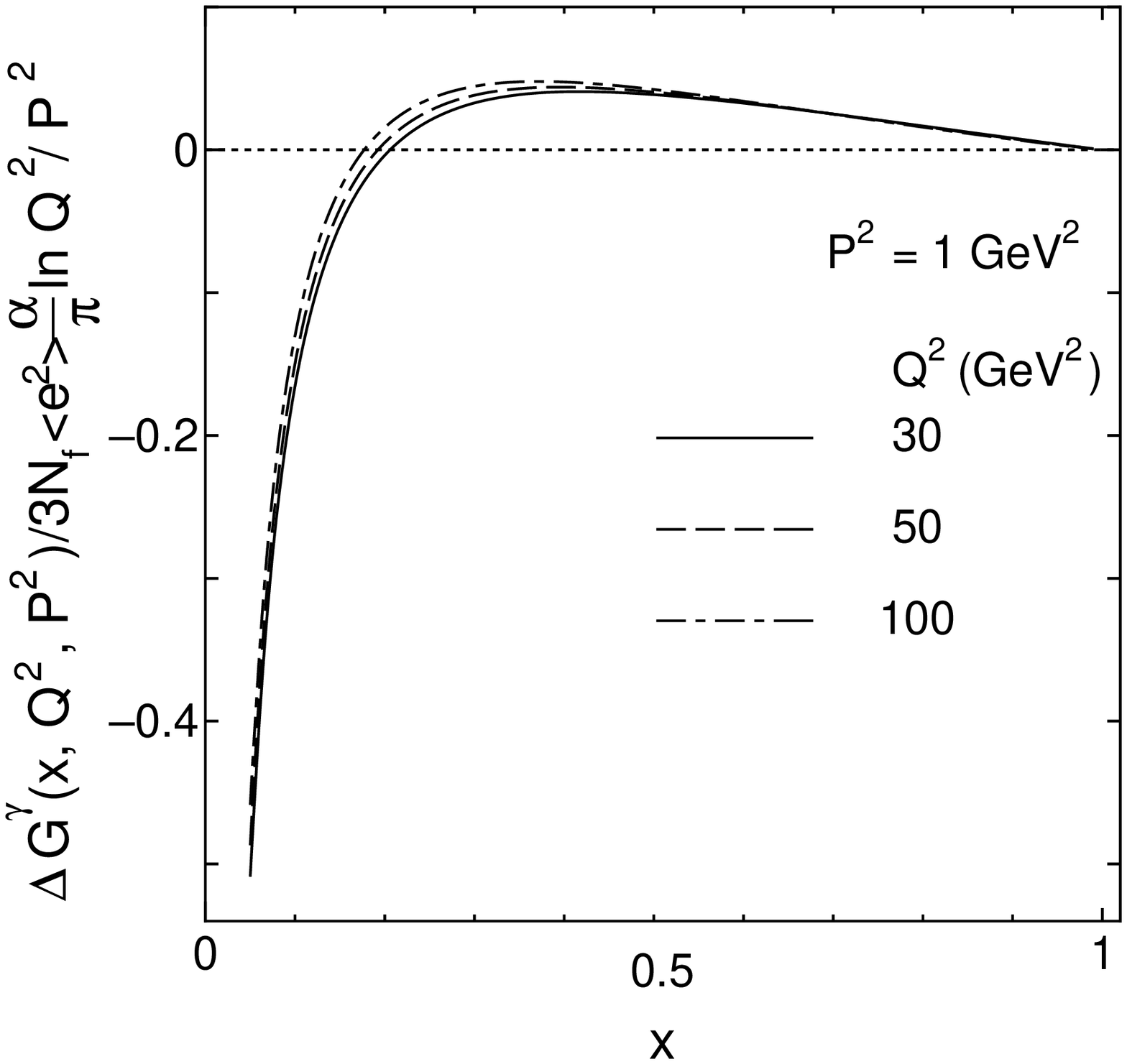}}
\vspace{-3.5cm}
\centerline{\large\bf Fig. 5}
\end{figure}

\newpage
\pagestyle{empty}
\input epsf.sty
\begin{figure}
\centerline{
\epsfxsize=16cm
\epsfbox{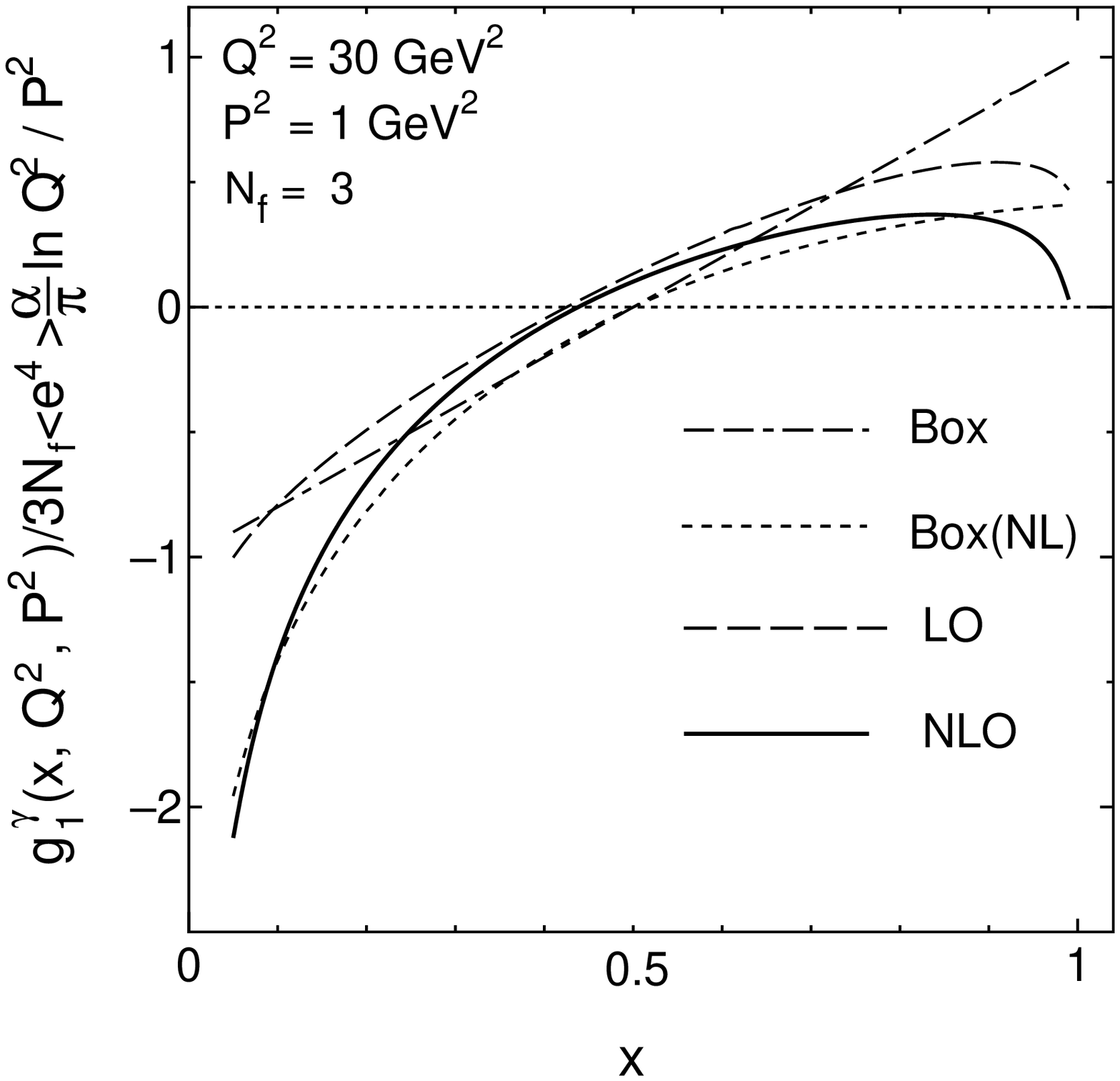}}
\vspace{-3.5cm}
\centerline{\large\bf Fig. 6}
\end{figure}

\newpage
\pagestyle{empty}
\input epsf.sty
\begin{figure}
\centerline{
\epsfxsize=16cm
\epsfbox{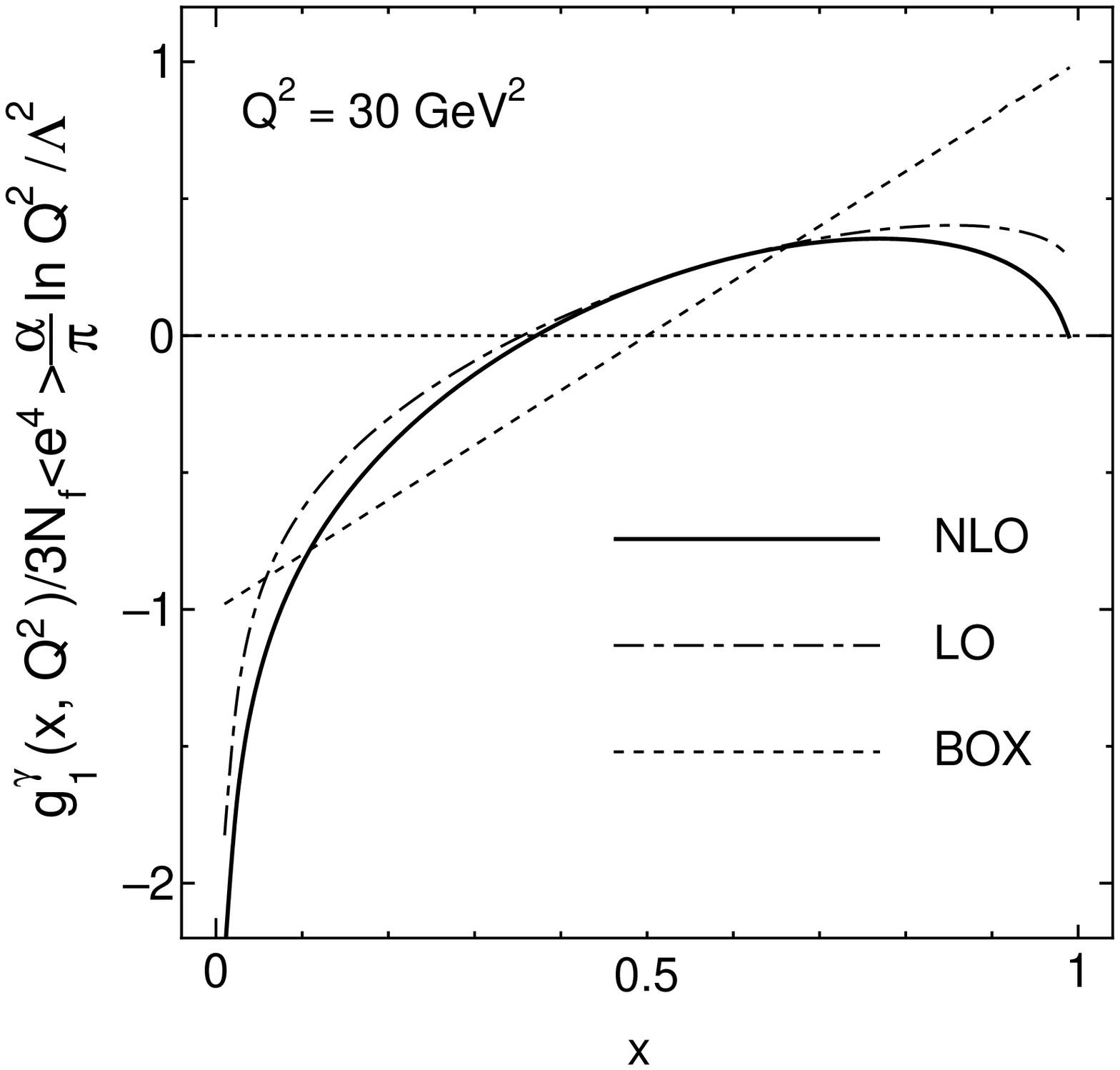}}
\vspace{-3.5cm}
\centerline{\large\bf Fig. 7}
\end{figure}



\begin{thebibliography}{99}
\bibitem{Footnote1}
We use the same notations as in Ref.\cite{SU}, except that
the symbol $\Delta$ has been appended to all the  spin-dependent
anomalous dimensions, coefficient functions and photon matrix elements.
\end{thebibliography}
\end{document}